\documentclass[twocolumn, switch]{article}

\usepackage{arxiv}

\usepackage[utf8]{inputenc} 
\usepackage[T1]{fontenc}    
\usepackage{hyperref}       
\usepackage{url}            
\usepackage{booktabs}       
\usepackage{amsfonts}       
\usepackage{nicefrac}       
\usepackage{microtype}      
\usepackage{lipsum}		

\usepackage{amsmath}
\usepackage{graphicx}
\usepackage{tabularx}
\usepackage{booktabs}
\usepackage{enumitem}
\usepackage{breakurl}
\usepackage{hyperref}
\usepackage{csquotes}
\usepackage{subcaption}
\usepackage{acronym}
\usepackage{nicefrac}
\usepackage{xcolor}

\newcolumntype{x}[1]{>{\centering\let\newline\\\arraybackslash\hspace{0pt}}p{#1}}

\newacro{OPF}[OPF]{optimal power flow}
\newacro{MES}[MES]{multi-energy system}
\newacro{OMEF}[OMEF]{optimal multi-energy flow}
\newacro{DER}[DER]{distributed energy resource}
\newacro{CPs}[CPs]{coupling points}
\newacro{CPES}[CPES]{cyber-physical energy system}
\newacro{CHP}[CHP]{combined heat and power}
\newacro{P2G}[P2G]{power to gas}
\newacro{P2H}[P2H]{power to heat}

\newcommand{\pressure}{p}

\newcommand{\nodea}{u}
\newcommand{\nodeb}{v}

\newcommand{\hloss}{H_{\rm loss}}

\newcommand{\tempin}{T_{\rm in}}
\newcommand{\tempout}{T_{\rm out}}
\newcommand{\hull}{M}
\newcommand{\cconstant}{C}

\newcommand{\cpdensity}{\rho^{\rm cp}}
\newcommand{\deploymentdensityenergy}{\rho^{\rm energy}}

{ \vspace{-0.15cm}%
    \small\noindent{\bfseries Availability of Data and Material:}\par%
    \noindent\ignorespaces}%
{ \par\noindent%
\ignorespacesafterend }%

\definecolor{elcolor}{RGB}{255, 160, 0}
\definecolor{gascolor}{RGB}{56, 142, 60}
\definecolor{heatcolor}{RGB}{211, 47, 47}

\AtBeginDocument{%
  }

\title{From Coupling to Resilience: Quantifying the Impact of Interconnection in Energy Carrier Grids}

\author{ \href{https://orcid.org/0000-0001-5339-6553}{\includegraphics[scale=0.06]{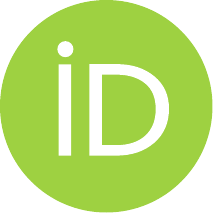}\hspace{1mm}Rico Schrage} \\
	Digitalized Energy Systems Group\\
    Carl von Ossietzky University of Oldenburg\\
    26129 Oldenburg, Germany\\
	\texttt{rico.schrage@uol.de} \\
	\And
	\href{https://orcid.org/0000-0003-1881-9172}{\includegraphics[scale=0.06]{orcid.pdf}\hspace{1mm}Astrid Nieße} \\
  	Digitalized Energy Systems Group\\
    Carl von Ossietzky University of Oldenburg\\
    26129 Oldenburg, Germany\\
}

\renewcommand{\shorttitle}{From Coupling to Resilience}

\hypersetup{
pdftitle={A template for the arxiv style},
pdfsubject={q-bio.NC, q-bio.QM},
pdfauthor={David S.~Hippocampus, Elias D.~Striatum},
pdfkeywords={First keyword, Second keyword, More},
}
\begin{document}

\twocolumn[\begin{@twocolumnfalse}
\maketitle

\begin{abstract}
Due to the increasing share of renewable energy resources and the emergence of couplings of different energy carrier grids, which may support the electricity networks by providing additional flexibility, conducting research on the properties of multi-energy systems is necessary. Primarily to keep stable grid operation and provide efficient planning, the resilience of such systems against low-probability, high-impact events is central. Previous steady-state resilience studies of electricity grids also involved investigating the topological attributes from a complex network theory perspective. However, this work aims to determine the influence of complex topological attributes on the resilience of coupled energy grids. To achieve this, we set up a Monte Carlo simulation to calculate the load-shedding performance indicator for the grids when affected by high-impact events. This indicator is used to calculate resilience metrics, which express the influence of the grids on each other. The metrics are the base to search for correlations between centrality/vitality metrics and the resilience impact metric. We apply our method to a case study based on a benchmark electricity grid. Our results show that, first, our impact metric is feasible for determining the influences of the network on each other. Second, we show that increasing coupling densities can lead to lower resilience in single-carrier grids. Third, it is apparent that centrality influences the impact of the grid components' resilience.
\end{abstract}


\vspace{0.5cm}

\end{@twocolumnfalse}]

\section{Introduction}
The resilience of energy grids is essential for system planning and operation. We understand resilience as \enquote{the capacity of the energy system and its components to cope with a hazardous event or trend, to respond in ways that maintain its essential functions, identity and structure as well as its capacity for adaptation, learning and transformation. It encompasses the following concepts: robustness, resourcefulness, recovery} \cite{IEAResilienceReport}, which is similar to common (multi-)energy grid resilience definitions \cite{yangResilienceAssessmentMethodologies2022}. Currently, coupled energy grids, including gas grids, district heating, and electricity grids, gain relevancy \cite{heatflexofferel, DecarbonizationResidentialHeating}. Further, the increase in the share of weather-dependent distributed energy resources (DER) in the electricity grid demands increased energy flexibility, which the other energy carrier grids can provide \cite{MANCARELLA20141}. However, operational strategies must change to make this possible, leading to more substantial influences of the grids on each other \cite{Shahidehpour2005mes, xu2017interactionmicroCHPS}. For planning and operational system research, it is interesting to determine the influence of the grids on each other and the influence of the individual \ac{CPs} on each other regarding the resilience of the grids. Further, finding metrics that can capture this grid behavior without needing a long-lasting simulation is advantageous for deriving operational and planning rules. 

Determining influences and deriving and calculating feasible metrics involve multiple challenges that shall be tackled. First, the coupled energy grids must be physically modeled, and different failure conditions must be sampled. Second, a coupled energy grid performance indicator \cite{bhusalPowerSystemResilience2020,yangResilienceAssessmentMethodologies2022} must describe the grid's qualitative state during the restricted performance time. The indicators can be used to quantify different components of resilience or to describe it aggregated. Third, based on these indicator metrics, resilience influences from groups, individual components, or whole grids on other components must be calculated. Fourth, as we want to find useable rules for network planning and operation, these calculation-intensive metrics need to be related to simpler (topological-based) metrics.

This paper presents possible solutions to all these challenges. We will define and calculate a resilience performance indicator metric of \ac{MES}. This metric will be calculated and modified for the overall system resilience, the influences of the systems on each other, and the individual influence on the resilience of \ac{CPs}. Further, complex network metrics and methodologies are applied to correlate the topological attributes to resilience. In the evaluation, we show the developed resilience impact metrics' validity and capability to capture the influence of grids on each other. Further, the results indicate the existence of correlations between topological attributes and individual resilience impact. It also reveals a necessity to care more about the risks and benefits of grid coupling regarding the resilience assessment. In short, this paper contains three contributions.
\begin{enumerate}
    \item A multi-energy system model which can simulate high-impact events
    \item Evaluated resilience metrics to quantify the impact of individual components or grids on each other
    \item A case study on differently coupled distribution systems
\end{enumerate}
The paper is structured as follows. First, the related work is analyzed, and implications are derived. After that, we will introduce the research questions and our perspective on resilience in \ac{MES}. This follows an introduction to our system model, including the steady-state equations for the couplings, the single carrier grids, and the theoretical graph representation. Then, the disturbance event generation is explained. Afterward, the resilience metrics, their performance indicator, and the load-shedding optimization, specifically for \ac{MES}, are introduced. As the next step, the Monte Carlo resilience simulation is described. This is followed by an introduction to the case study, the experiments, and its required parameterization. Then, the experimental resilience analysis results are presented and discussed in detail. 
\section{Related Work}
There is much research about energy grid resilience \cite{JASIUNAS2021111476, yangResilienceAssessmentMethodologies2022, yodoResilienceAssessmentInterdependent2021}. Some papers focus on improving resilience using new network planning and operation strategies, and others are about quantifying different aspects of resilience considering specific high-impact, low-probability events (e.g., \cite{baoModelingEvaluatingNodal2020}).

While most work has been done in the area of electricity grid resilience (see, e.g., \cite{JASIUNAS2021111476, bhusalPowerSystemResilience2020}), multi-energy grid resilience arose as a new topic in the last years. We categorize this work into three different groups:
\begin{enumerate}
    \item Measuring resilience
    \item Resilience seen from complex network theory
    \item Resilience improvement strategies
\end{enumerate}
\paragraph{(1) Measuring resilience} For resilience measuring, Yang et al. \cite{yangResilienceAssessmentMethodologies2022} reviews power system assessment methods and defines \ac{MES} resilience. They also discuss methods for resilience improvements and include state-of-the-art resilience metrics in multi-energy cyber-physical systems. One specific possibility to measure resilience is shown in \cite{baoModelingEvaluatingNodal2020}, where the authors defined a multi-energy load-shedding metric. However, the listed publication lacks a differentiated analysis of the influence of couplings in MES on resilience.
\paragraph{(2) Resilience seen from complex network theory} In this area, electricity grid research is dominant and conducted with a complex systems point of view. The authors of \cite{paganiPowerGridComplex2013} present a comprehensive overview of complex network power grid research. This review included research on electricity grids with 30 to 31400 nodes with a mean of $~4800$ and a median of $2100$ nodes. Further, much research focuses on reliability analysis \cite{afzalStateoftheartReviewPower2020} using metrics like betweenness centrality \cite{freemanSetMeasuresCentrality1977} or a degree distribution. Therefore, network construction is of great importance. Complex network methods are helpful, especially for designing and analyzing resilient power grids. For example, the Barabási-Albert scale-free network model \cite{barabasiEmergenceScalingRandom1999} has been applied to the North American electric grid \cite{chassinEvaluatingNorthAmerican2005}. The authors were able to confirm the accuracy of the model in terms of predicted reliability.
\paragraph{(3) Resilience improvement strategies} In \cite{yodoResilienceAssessmentInterdependent2021}, the authors evaluate the effect of introducing microgrids on the resilience of the \ac{MES}. \cite{sunResilienceEnhancementStrategy2022} proposes a multi-stage recovery algorithm for \ac{MES} after an extreme weather event. A demand curtailment performance metric is used to assess the recovery. Finally, the authors of \cite{gharehveranTwostageResilienceconstrainedPlanning2022} show a possibility of multi-energy planning with resilience constraints.
\paragraph{Research Gap} Neither of the approaches mentioned above tackles measuring the impact of the grids on each other and the impact of the single nodes in the system nor is there any publication about finding resilience heuristics with complex network theory. However, we build upon the idea of resilience measurement based on the multi-energy load-shedding and the ideas for using complex network metrics from the power system resilience research. However, to the best of our knowledge, measuring the resilience dependencies of the single network carrier networks coupled to multi-energy systems (MES) has not been considered yet. Using complex network methods as tools to describe \ac{MES} resilience is also a novel approach.
\paragraph{Research Focus} This work focuses on resilience from a steady-state perspective of a system of energy carrier grids. We understand resilience as \enquote{the ability to adapt to changing conditions and withstand and rapidly recover from disruption} \cite{homelandResilienceReport}. Specifically, the influence of the resilience of these different carrier grids on each other is investigated. The main research questions, which this work will answer, are defined as follows.
\begin{enumerate}
    \item Do the grids influence each other's resilience, and if so, what is the impact of the coupling density on this effect?
    \item Can couplings between the grids increase or decrease resilience (considering no specific island-building strategies)?
    \item Is there a relation between complex topology attributes and the resilience/resilience impact?
\end{enumerate}
We present a steady-state model of the different carrier grids (electricity, gas, and heating) to answer these questions. The different generators and loads are not modeled individually. Further, we assume complete knowledge of all grid-related parameters (topology, line lengths, etc.) and the power capabilities/demands of the different generators and consumers. It is also assumed that all generators and loads are controllable if necessary.
\section{System Model}\label{sec:system_model}
Some prerequisites for analyzing multi-energy systems' resilience and complex graph attributes must exist. First, it is necessary to model the physical steady-state behavior of the gas, heat, and electricity grid to calculate the system state and to find feasible states under component failures. Second, the \ac{CPs} need to be modeled. At last, the system must be represented as a graph, and edge weights must be assigned. This is necessary to calculate the complex network metrics. In the following, we explain these steps in detail.
\subsection{Calculating Energy Flows}
The electricity grid is modeled using the well-known steady-state AC power equations, which are equivalently used in, e.g., PowerModels \cite{8442948} and, therefore, omitted at this point. The interested reader can find the equations in the appendix \ref{app:ac_equations} for convenience.

To model the gas grid, the Weymouth equation \cite{schroeder2010tutorial, bent2022gasmodels} calculates the pressures on each junction.
\begin{equation}\label{eq:first_system_eq}
    \pressure_a - \pressure_b = \left(\frac{\lambda L}{D}\right)\left(\frac{\Gamma^2}{A^2}\right)\cdot f|f| \text{ with } \Gamma^2=\frac{ZRT}{m}
\end{equation}
This equation describes the pressure drop in a cylindrical pipe induced between junction $a$ and junction $b$. Here $\pressure$ is the pressure in Pascal, $\lambda$ is the dimensionless friction factor, $L$ is the length of the pipe, $D$ is the diameter of the pipe, $A$ is the cross-sectional area of the pipe ($A=\nicefrac{\pi D^2}{4}$), $f$ is the mass flow, $Z$ is the gas compressibility factor, $R$ is the universal gas constant, and $m$ is the molar mass.

The friction $\lambda$ will be calculated using the Reynolds number and the well-known Prandtl-Nikuradse formula.
\begin{equation}
    \begin{split}
        \lambda &= \frac{64}{Re} + PrNi  \\
        \textit{with } Re &= \frac{|f|D}{\eta A} \\
        \textit{and } PrNi &= \frac{1}{(2\log(3.71\nicefrac{D_{\rm in}}{\epsilon}))^2}
    \end{split}
\end{equation}
Here, $Re$ is the Reynolds number, $PrNi$ is the Prandtl-Nikurdse friction, $\eta$ is the dynamic viscosity, $D_{\rm in}$ is the inner diameter of the pipe, and $\epsilon$ is the roughness of the pipe.

It is also necessary to formulate the mass balance equation for each junction $a$ to ensure mass conservation.
\begin{equation}
    \sum_{b=1}^{b\leq n}{f_{\rm in,b}} + \sum_{b=1}^{b\leq m}{f_{\rm out,b}} = 0
\end{equation}
Here, $n$ is the number of incoming mass flows, and $m$ is the number of outgoing mass flows, $f_{\rm in,b}$ is the $b$'th incoming mass flow, and $f_{\rm out,b}$ is the $b$'th outgoing mass flow.

Modeling the hydraulics of the water grid is similar to the gas grid. However, here we use the Darcy-Weisbach equation \cite{howell1967aerospace}, while the friction and the mass conservation equation remain as defined above.
\begin{equation}
    \pressure_a - \pressure_b = \lambda \left(\frac{L\rho}{2}\right) \left(\frac{-f|f|}{D}\right)
\end{equation}
In this equation, $\rho$ is the density of the fluid (in this case water).

Further, we need to model the heat losses in the water grid. We will use the conductive heat transfer equation \cite{lienhard2008doe}.
\begin{equation}\label{eq:last_system_eq}
    H_{\rm loss} = -2\pi k L \cdot f \left(\frac{T_a + T_b }{2}- T_{\rm ext}\right)
\end{equation}
Here, $H_{\rm loss}$ is the heat energy loss, $k$ is the insulation transfer coefficient, $T_a$ is the temperature at junction $a$, $T_b$ is the temperature at junction $b$, and $T_ext$ is the temperature of the external environment.
\subsection{\ac{CPs}}\label{subsec:multi_energy_system}
Three \ac{CPs} are used in the simulation: \ac{CHP}, \ac{P2H}, and \ac{P2G}, which refers to hydrogen electrolysis. The \ac{CHP} is described using one demand node in the gas system, one producer node in the electricity system, and a heat exchanger connected to two nodes in the heating system. The amounts are calculated using the following equations.
\begin{equation}
    \begin{split}
        P_{\rm el} &= \eta_{\rm el} f_{\rm gas} \cdot 3.6 \cdot {\rm HHV}\\
        H_{\rm he} &= -\eta_{\rm heat} f_{\rm gas} \cdot \frac{3.6}{{\rm HHV}}
    \end{split}
\end{equation}
Here, $P_{\rm el}$ is the electric energy injected into the electricity grid, and $\eta_{\rm el}$ is the electricity conversion efficiency, $H_{\rm he}$ is the heating energy injected into the heating grid, $\eta_{\rm heat}$ is the heat conversion efficiency, $f_{\rm gas}$ is the gas demand, and HHV is the higher heating value of the gas.

The \ac{P2H} is described using the following equation.
\begin{equation}
    H_{\rm he} = \eta_{\rm el} P_{\rm demand}
\end{equation}
Here, $P_{\rm demand}$ is the electrical demand. Finally, the Power-to-gas component is modeled with the following equation.
\begin{equation}
    f_{\rm gas} = \frac{\eta_{\rm gas} P_{\rm el}}{3.6 \cdot {\rm HHV}}
\end{equation}
\subsection{Grid Data}\label{subsec:grid_data}
As there is insufficient grid data for coupled multi-energy networks, especially with the structure we assume to be important in the future, we will explain how our grid data is generated. As a basis, power grid data from \textit{simbench} \cite{meinecke2020simbench} is used. \textit{Simbench} datasets contain fully featured power grids and appropriate time-series data. It is a benchmark dataset for novel network planning and operation methods. The gas and heat networks are generated using an appropriate \textit{simbench} network. Considering medium to low-voltage networks, it is assumed that every power node could eventually be a heat and gas node. As more households and industries are connected to the power grid relative to the gas or heat network, deployment rates have been chosen. The heat and gas networks are generated along the power network nodes. As a result, we have a heat and gas network with a similar topology but a smaller deployment density $\deploymentdensityenergy$ of productive nodes. Further, after generating these networks, \ac{CPs} will be generated using constant coupling point densities $\cpdensity$ for every CP-type: \ac{P2G}, \ac{CHP}, and \ac{P2H}.
\subsection{Graph Representation}\label{subsec:graph_representation}
This section presents the methodical approach for defining the network topology using a multi-energy network. The model is based on the \ac{MES} graph definition in \cite{schrage2023influence}.

We consider the grid topology as directed weighted multigraph $G = (V,\,E)$ with $e=(\nodea,\,\nodeb,\,z,\,\psi,\,\omega) \in E,\, \nodea,\nodeb\in V$ with $V\subset\mathbb{N}\times\mathbb{N}$. An edge $e$ is a directed link between two nodes $\nodea$ and $\nodeb$. The characteristic $\psi$ describes the edge type, $\omega$ is the edge weight, and $z$ is the edge id to enable multiple edges between two nodes. A node $u=(\kappa,\,\beta)$ is a pair of natural numbers. The first is an identifier of an actual network unit, and the second identifies the energy carrier network to which it belongs. A coupling point can be modeled as two nodes with different network affiliations and a directed edge between them. 
Every producing or consuming energy unit will be represented by a node. Nodes are connected if there is a direct physical connection (a pipe, line, etc.). If two nodes $\nodea,\,\nodeb \in V$ are in the same energy carrier network, there are two edges $(\nodea,\,\nodeb,\,\omega,\,\psi)$ and $(\nodeb,\,\nodea,\,\omega,\,\psi)$. There is precisely one direction between nodes of different networks, depending on the type of the coupling point, represented by $\psi$. As this graph represents the physical topology, buses and junctions are represented by nodes, while the generators and demands are assigned to these nodes. 
The weight $\omega$ of an edge $e\in E$ is defined, depending on the carrier network affiliation. The weights of the \ac{CPs}' branches will be set to $1 - \eta$, with $\eta$ being its relevant loss constant. For the electricity carrier grid, the weight is defined as the relative power loss for edges in the power network.
\begin{equation}
    \eta_{\rm power} = \frac{P_{\rm in} - P_{\rm out}}{P_{\rm in}}
\end{equation}
\begin{equation}
    \eta_{\rm gas} = \frac{p_{\rm i} - p_{\rm j}}{p_{\rm i}}
\end{equation}
At last, the water pipes' edge weight is determined using heat loss.
\begin{equation}
    \begin{split}
        \eta_{\rm heat} &= \frac{\hloss}{\hull \cconstant \cdot |\tempin - \tempout|} \\ 
    \end{split}
\end{equation}
Here, $\tempin$ and $\tempout$ are the temperatures at the pipe's start and end, $\hull$ is the volume of water in the pipeline, and $C$ is the specific heat capacity.
\section{Resilience in MES}
This part aims to, first, measure resilience in a \ac{MES} and second, find applications of complex network theory to estimate these resilience measurements. The result of the second task will enable the system's evaluation and provide methods to improve the system with efficient and resilient assessments. 

In the following sections, high-impact events and their types are first introduced; second, the event generation will be introduced; third, the network performance metric will be shown; finally, the resilience impact metrics will be presented.
\subsection{High-Impact Events}
Different types of events can cause failures in the energy system. We generally consider only external causes. For these, we found two different types, based on the causes and effects described in \cite{ResilienceCausesFanidhar}.
\begin{enumerate}
    \item Statically random events (e.g., accidents or manipulations)
    \item Dynamically random events (e.g., hurricanes or earthquakes)
\end{enumerate}
Generally, we assume that the first type of failure can be modeled homogeneously. For instance, in cases where the probability of failure is caused by many nearly arbitrarily modifiable circumstances such as infrastructure or the distance to critical consumers, we assume that these can be modeled with static individual failure probabilities using a uniform random distribution. 

However, the second type is more complex and requires a different approach. These events are not purely random and are considerably influenced by factors such as spatial network attributes. Additionally, some natural disasters affect different components in different ways, e.g. overhead lines being more susceptible to damage from hurricanes than underground ones. Furthermore, these events change over time and do not follow a uniform random distribution.
\subsection{Event Generation}
To measure the systems' resilience, high-impact events will be generated, representing some disturbance in the system. An event is defined as follows.
\begin{equation}
    \begin{split}
    e &= \{\theta_1,\,\dots,\,\theta_i,\,\dots,\,\theta_n\}\\
    &\text{with } n \in \mathbb{N} \text{ and } 0 < i\in\mathbb{N} \leq n
    \end{split}
\end{equation}
Here, an event $e$ consists of $n$ times $\theta$, vectors describing every grid component's health. The variable $n$ is the number of time steps the event simulates. Further, $E$ shall be the number of all generated events in one simulation.
\begin{equation}
    \theta_i = 
    \begin{pmatrix}
    \phi_{i,\,0}\\\dots\\\phi_{i,\,j}\\\dots\\\phi_{i,\,m}
    \end{pmatrix}
    \text{with } m \in \mathbb{N} \text{ and } 0 < j\in\mathbb{N} \leq m
\end{equation}
The symbol $\phi_{i,\,j}$ describes the vitality of the network component $j$. In this work it can be assigned to $1$, $0$, or $2$, representing \textit{fully-functional}, \textit{broken}, and \textit{repaired}. It is calculated based on the time, location, type of component, and the grid it is part of.
\begin{equation}
    p^\phi_{i,\,j} = \rho_j \cdot \gamma_j \cdot \beta_{i,\,j} \cdot p_G^{\rm grid} \cdot p^{\rm base}\\
\end{equation}
Here, $p^\phi_{i,\,j}$ is the failure probability of $j$ at $i$, $\rho_j$ is the failure probability coefficient based on the type of the component $j$, $\gamma_j$ is the probability coefficient based on the individual component $j$ and $\beta_{i,\,j}$ is based on the spatial position of the component $j$ at the time-step $i$, $p_G^{\rm grid}$ is the base probability of a failure in the grid $G$, and $p^{\rm base}$ is the baseline failure probability. Then, based on the probability calculated for the components $j$ at time-step $i$, it is assigned whether the component is vital. Further, it has to be calculated how long the component is broken.
\begin{equation}\label{eq:failure_calc_ij}
    \begin{split}
    \phi_{i,\,j} = 
    \begin{cases}
        0, & \text{if } r < p^\phi_{i,\,j}\\
        2, & \text{if } \phi_{i-1,\,j} = 0 \text{ and } p^{\phi,\,\rm initial}_{\rm i,\,j} \geq r\\
        0, & \text{if } \phi_{i-1,\,j} = 0\\
        1  & \text{else}
    \end{cases}
    \end{split}
\end{equation}
In this, $r$ is a uniformly distributed random number between zero and one, $p^{\phi,\,\rm initial}_{\rm i,\,j}$ is the initial probability for a currently broken component.
\subsection{System Performance}\label{subsec:system_performance}
Under optimal management, the system will be optimized in every step to simulate the event and calculate the energy network's performance. This creates a baseline resilience, which could be safely compared to developed operation strategies. We assume that optimal management can happen without time delays; therefore, this work does not consider cascading effects.

To calculate this performance, the whole grid with every part of the system, heating, gas, and electricity, has been described with its steady-state equations according to equations (\ref{eq:first_system_eq})-(\ref{eq:last_system_eq}). Further, the energy conversion equations have to be formulated. This will result in a non-linear equation system, which can, first, be solved directly using NLP-solver, metaheuristics, or learning-based approaches. Here, we used an NLP-solver.

The following equation optimization problem can be used to calculate the load-shedding performance of the system. The objective for load-shedding can be expressed as follows.
\begin{equation}
    \begin{split}
    LS=min \sum_{d\in D^{\rm heat}} S^{\rm heat}_d P^{\rm heat}_d &+ \sum_{d\in D^{\rm gas}} S^{\rm gas}_d P^{\rm gas}_d \\
    &+ \sum_{d\in D^{\rm el}} S^{\rm el}_d P^{\rm el}_d
    \end{split}
\end{equation}
Here, $D^{\rm sector}$ is the set of loads in the mentioned \textit{sector}, $S^{\rm sector}_d$ is the shedding coefficient of $d$, $P^{\rm sector}_d$ is the demands power consume of $d$. Further, we define the following constraints.
\begin{equation}
    \begin{split}
    \forall i\in B_{\rm el}&\colon v_{\rm min} \leq |V_i| \leq v_{\rm max} \\
    \forall j\in J_{\rm gas}&\colon p_{\rm min} \leq p_j \leq p_{\rm max} \\
    \forall l\in J_{\rm heat}&\colon T_{\rm min} \leq T_l \leq T_{\rm max} \\
    \forall e\in E_{\rm el}&\colon lp_{\rm min} \leq lp_e \leq lp_{\rm max} \\
    P_{\rm slack} &= 0 \\
    M_{\rm slack,\,gas} &= 0
    \end{split}
\end{equation}
Here, $|V_i|$ is the voltage magnitude at the bus $i$ with the lower and upper bounds $v_{\rm min}$ and $v_{\rm max}$, $p_j$ is the pressure at the gas junction $j$ with the lower and upper bounds $p_{\rm min}$ and $p_{\rm max}$, $T_l$ is the temperature at the water junction $l$ with the lower and upper bounds $T_{\rm min}$ and $T_{\rm max}$, $lp_e$ is the thermal loading percent of the power line $e$ with the lower and upper bounds $lp_{\rm min}$ and $lp_{\rm max}$, $P_{\rm slack}$ is the power at the slack node of the electric grid, and $M_{\rm slack,\, gas}$ is the mass flow at the slack node of the gas grid. Further, $B_{\rm el}$ is the set of valid bus identifiers, $J_{\rm gas}$ is the set of valid gas junction identifiers, $J_{\rm heat}$ is the set of valid water junction identifier, and $E_{\rm el}$ is the set of valid electrical line identifier.
\subsection{Resilience Metrics}\label{subsec:resilience_metrics}
Three different types of metrics are relevant. The first is a general resilience metric for evaluating the overall effect of the events; the second enables analyzing the impact of the different carrier networks on each other. Third, a metric is needed to break resilience into single component classes in a complex network. In the following paragraph, the metric for each category is presented.

\paragraph{Overall Metric} These can be derived from the power system literature, in which several metrics are introduced based on performance measurement like load shedding \cite{bhusalPowerSystemResilience2020}. As every energy grid has the same objective to deliver energy in some form to customers, metrics already established for power grids can also be used for the coupled grid. However, there are some differences when choosing the correct ones. Unlike the electric grid, energy flows in gas and heating networks are very slow, and demands can be more flexible. Different metrics for different carrier networks can be relevant, and it is not generally possible to drop every differentiation between the carriers. Here, we will use the sum of the curtailed demand over time.
\begin{equation}\label{eq:resilience_overall}
    R^G_{LS} = \frac{1}{|E|} \sum_{e\in E}\sum_{i=0}^{i<n} LS^G_{i,\,e}
\end{equation}
Here, $i$ is the timestep, $e$ is the event with $E$ as the set of all events, and $LS^G_{i,\,e}$ is the load-shedding in grid $G$ at interval $i$ in event $e$. 
\paragraph{Single Component Impact} To evaluate the performance of each component, we can compare its performance when it was broken to when it was functioning correctly. Based on this concept, we have formulated a metric denoted as $SCI^G_j$, where $j$ represents the component and $G$ represents the grid on which the impact is being measured. Note that we do not analyze cross-correlations influencing this metric between components. This type of analysis can be part of future work; in this paper, it is not considered as these effects should be neglectable due to the nature of the Monte Carlo simulation described in the next section. 
\begin{equation}\label{eq:impact_ind}
    \begin{split}
        SCI^G_j &= \begin{cases}
                        \frac{SCI^G_{j,\rm in}}{SCI^G_{j,\rm nin}} & \text{if } SCI^G_{j,\rm in} > SCI^G_{j,\rm nin}\\
                        -\frac{SCI^G_{j,\rm nin}}{SCI^G_{j,\rm in}} & \text{else}\\
                    \end{cases}  \\
        SCI^G_{j,\rm in} &= \frac{1}{|E_j|} \sum_{e\in E_j}\sum_{i\in e^t_j} LS^G_{i,\,e}\\
        SCI^G_{j,\rm notin} &= \frac{1}{|E|} \sum_{e\in E}\sum_{i\notin e^t_j} LS^G_{i,\,e}
    \end{split}
\end{equation}
\paragraph{Carrier Grid Impact} One option to calculate the impact could be by calculating the correlation between the performances of the grids over time using, e.g., the cross-correlation function with pre-whitening \cite{yule1926we}. However, we will set up the simulation in a way that generates failures for single carrier grids to determine an unambiguous impact, as the correlation metric is error-prone. We define the overall impact of the carrier grid $G_1$ on carrier grid $G_2$ as follows.
\begin{equation}\label{eq:impact_grid}
    SCI^{G_1,\,G_2} = \sum_{j\in C_{G_1}} SCI^{G_2}_j
\end{equation}
Here, $C_{G_X}$ is the set of all components in grid $G_X$.
\section{Resilience Simulation}
\begin{figure}
    \centering
    \includegraphics[width=0.5\textwidth]{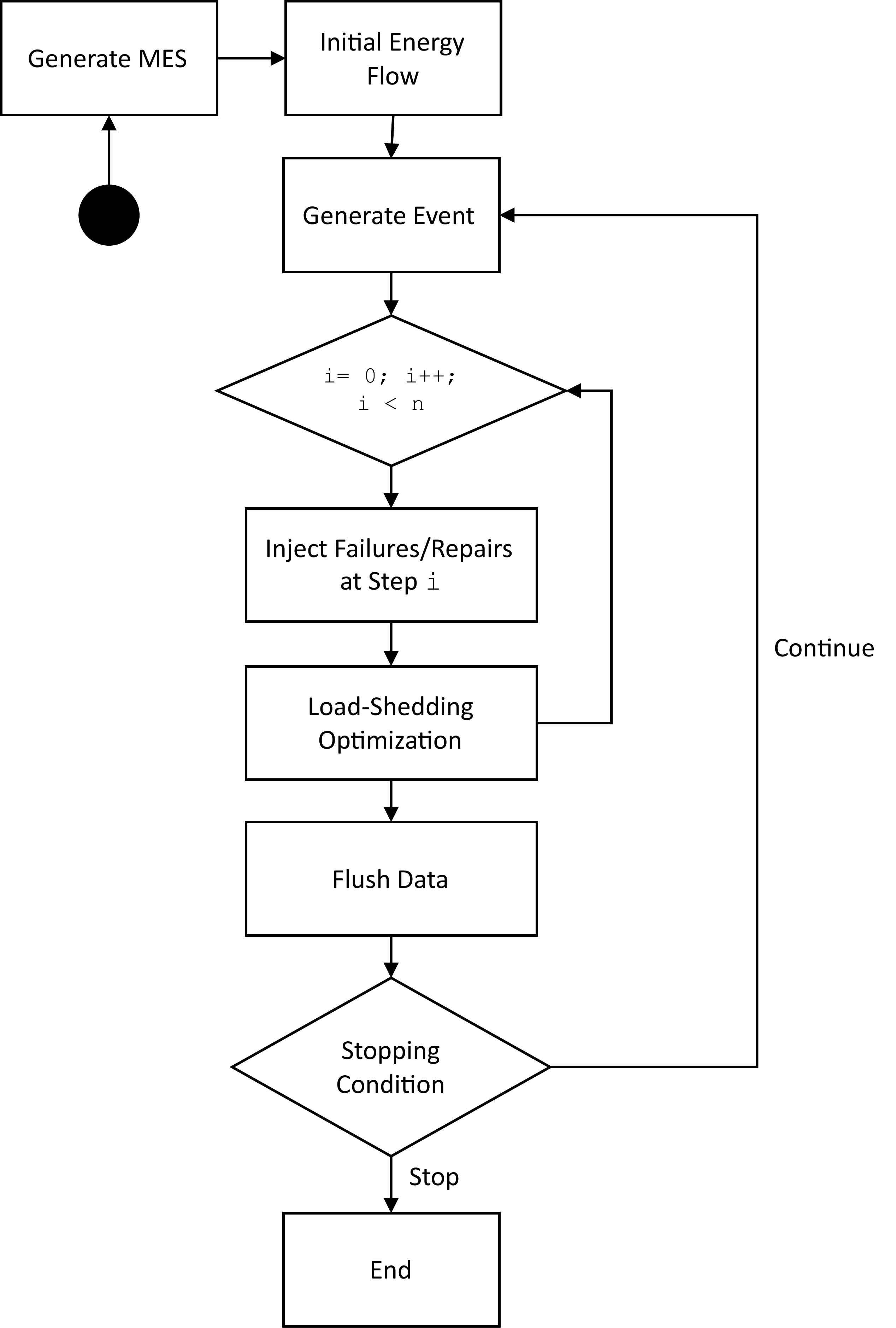}
    \caption{Flow of the Monte-Carlo simulation for calculating the resilience of the \ac{MES}}\label{fig:resilience_simulation_flow}
\end{figure}
Up to this section, we described the calculation of the energy flows, the generation of high-impact events, the optimization of the network for minimal load-shedding, the calculation of the performance indicator, and the calculation of the performance metrics. This section will describe how all these steps are put together in the resilience simulation of the networks. See Figure \ref{fig:resilience_simulation_flow} for a visual description of the execution flow.

The first step in this simulation is to prepare the multi-energy network according to \ref{subsec:multi_energy_system} using the set CP-density coefficient. For the baseline state of the network and to prove the generated \ac{MES} feasible, the initial multi-energy flow is calculated using the equations and methods described in section \ref{sec:system_model}. 

After that, the actual Monte-Carlo simulation starts. The following steps are executed in this loop until the stopping condition is met. We use a Kalman filter configured as described by Marti et al. \cite{marti2016stopping} to determine the change stability on the impact metric as a stopping condition. Further, at least 1,000 events must be executed to prevent premature termination.
\begin{enumerate}
    \item Generate a random medium to high-impact event.
    \item Simulate the event for $n$ time-steps, while $i$ is the current step number.
    \begin{enumerate}
        \item Modify the network concerning all generated failure or repair events at step $i$. Note that disconnected parts of the single carrier networks are no longer included in the calculations and will be considered completely dropped.
        \item Calculate the load shedding while optimizing for a minimal result (see \ref{subsec:system_performance})
    \end{enumerate}
    \item Write down the results.
    \item Check the stopping condition; go back to step 1 if it is not met yet.
\end{enumerate}
When the stopping condition is met, the simulations will stop, and all further resilience metrics (\ref{subsec:resilience_metrics}) and evaluation metrics (next chapter) are calculated.
\section{Evaluation}
To answer our research questions, we aim to apply the described method to specific grid instances designed for rural areas and generated with different coupling densities. Further, different scenarios are executed in which the focus of the high-impact events is artificially changed. This enables us to validate the metrics we defined before. In the last step of the evaluation, we will look at the topological attributes of the generated networks and relate them to the calculated resilience metrics.
\subsection{Experiment Design}
This paper evaluates one base network with five different coupling point density setups (0, 0.5, 1, 1.5 2) with six different event generation parameter sets (see Table \ref{tab:event_gen_impact_param_set}). This leads to 30 different parameter sets executed in resilience simulations. The impact parameters are chosen such that we can investigate the impact of failures in single-carrier networks on other networks without having to consider crossover effects. Further, we want to look at the impact of different event strengths. The coupling point densities are chosen such that we have a network with relatively low energy conversion, medium energy conversion, and high energy conversion capacity regarding the overall installed capacities in the networks.
\begin{table}
    \centering
    \begin{tabular}{ccc}
        \toprule
         \textbf{Event characteristic} & \textbf{$p_{\rm el}^{\rm grid},p_{\rm heat}^{\rm grid},p_{\rm gas}^{\rm grid},p_{\rm cp}^{\rm grid}$}& \\
         \midrule
         High electricity&  3, 0, 0, 0& \\
         High heating&  0, 3, 0, 0& \\
         High gas&  0, 0, 3, 0& \\
         High cp&  0, 0, 0, 3& \\
         Low overall&  1, 1, 1, 1& \\
         Medium overall&  2, 2, 2, 2& \\
         \bottomrule
    \end{tabular}
    \caption{Event generation impact parameter sets}
    \label{tab:event_gen_impact_param_set}
\end{table}
\subsection{Complex Network Metrics}
We use several complex network metrics to answer whether the complex topological attributes of single nodes, especially centrality, are essential for the node's participation in the overall resilience of the coupled network. Specifically, these metrics have been chosen due to their popularity in complex network theory \cite{estrada2012structure} and usage (and lack of) in electricity grid research \cite{paganiPowerGridComplex2013}. Further, we choose the closeness vitality and Katz-centrality metrics for comparison and their ability to capture different topological aspects. The metrics are listed in the following.
\begin{figure*}
    \centering
    \centering
    \includegraphics[width=1\textwidth]{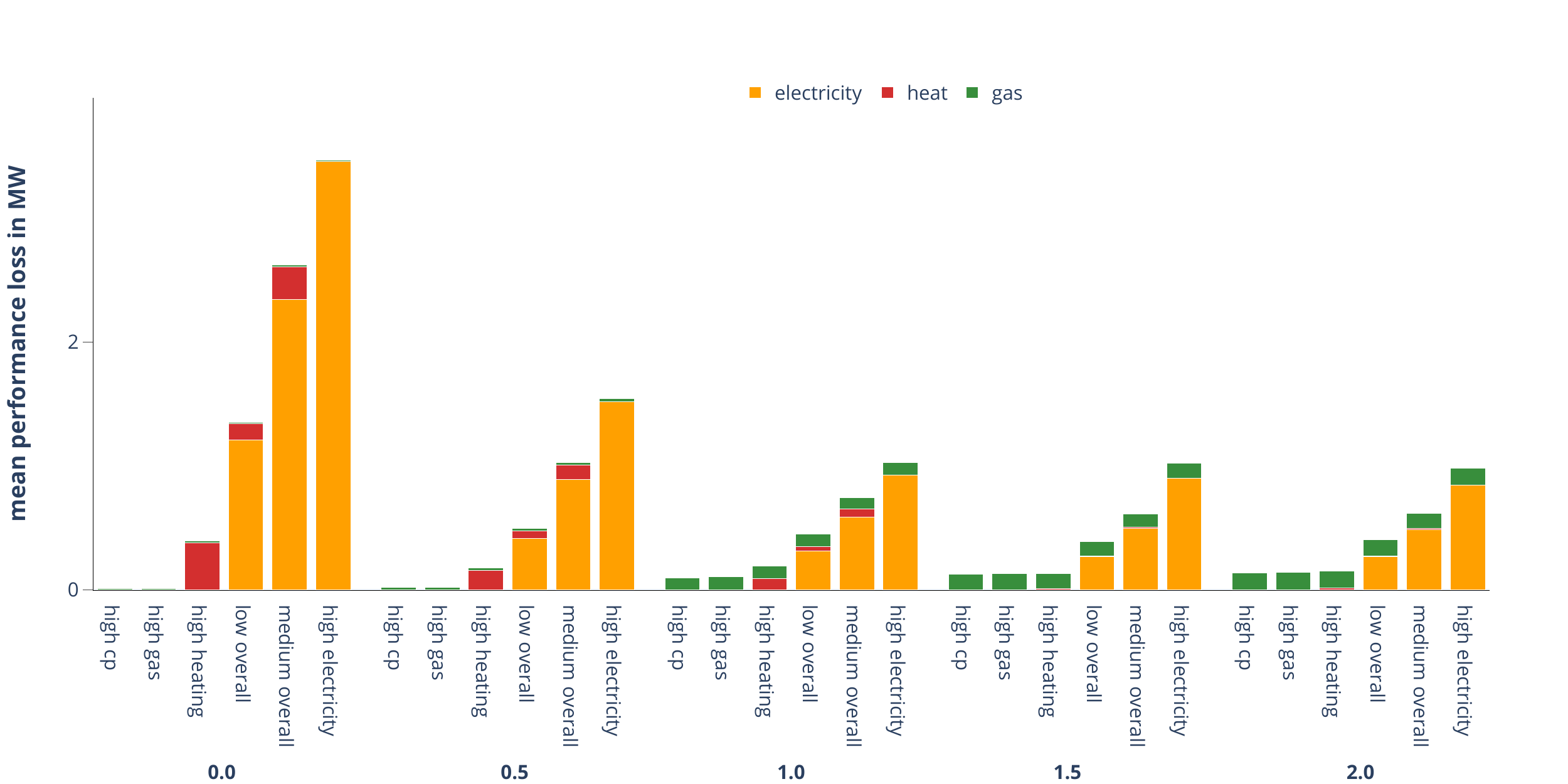}
    \caption{Performance drop by scenario and carrier; x-axis single labels: event characteristic; x-axis group labels: CP density}\label{fig:perf_drop_carrier}
\end{figure*}
\paragraph{Node betweenness centrality} Measuring centrality is essential in energy systems, as a removed high centrality node can lead to a significant lack of short paths in the system, which also influences the neighborhood functions of the nodes. The betweenness centrality \cite{freemanSetMeasuresCentrality1977} defines centrality via the number of shortest paths passing through a node $z$. We calculate the shortest path using the edge weights described in section \ref{subsec:graph_representation}, as the line and pipe distances cannot describe the actual energy distance. Formally, it can be defined as follows: 
\begin{equation}
   c_{\rm betweenness}^{\rm node}(z) =\sum_{\nodea,\,\nodeb \in V} \frac{\gamma(\nodea,\,\nodeb\,|\,z)}{\gamma(\nodea,\,\nodeb)}.
\end{equation}
Here, $\gamma(\nodea,\,\nodeb)$ is the shortest path from $\nodea$ to $\nodeb$ while $\gamma(\nodea,\,\nodeb\,|\,z)$ is the shortest path from $\nodea$ to $\nodeb$ passing through $z$.
\paragraph{Edge betweenness centrality} The coupling point itself is described as a link between two or more networks; consequently, we also want to describe the centrality of edges. For this purpose, the edge betweenness centrality will also be used \cite{BRANDES2008136}.
\begin{equation}
   c_{\rm betweenness}^{\rm edge}(e) =\sum_{\nodea,\,\nodeb \in V} \frac{\gamma(\nodea,\,\nodeb\,|\,e)}{\gamma(\nodea,\,\nodeb)}
\end{equation}
This equation is nearly identical to the node betweenness centrality with the difference that $\gamma(\nodea,\,\nodeb\,|\,e)$ requires the path to pass through an edge $e$ rather than a node.
\paragraph{Degree} The degree of the node is the number of all connected edges. We extend this common definition to edges and define the degree of an edge as the sum of the degrees of its connected nodes.
\paragraph{Closeness vitality} The closeness vitality \cite{brandes2005network} describes the positional attribute of a node for the participation force of single nodes. It is defined for a node $v$ as the change of the sum of distances between all node pairs, which do not include $v$. Formally, it is defined by the authors of \cite{brandes2005network} as
\begin{equation}
    c_{\rm vitality}(v) = I_W(G) - I_W(G\setminus\{v\}).
\end{equation}
$I_W$ is the Wiener Index \cite{ja01193a005} of a graph $G$. The Wiener Index is the sum of distances between all node pairs.
\paragraph{Katz centrality} The Katz centrality (KC) \cite{katz1953new} of a specific node is defined by the centrality of its neighbor. It is defined as follows.
\begin{equation}
    c_{\rm katz}(v) = \alpha\sum_u A_{v,u}x_u+\beta
\end{equation}
Here, A is the adjacency matrix, $\alpha$ the attenuation, and $\beta$ the initial centrality.
\\\\
In the following chapter on results, we write about betweenness centrality for all variants. It is meant to refer to the group betweenness in the case of \ac{CHP} and \ac{P2H}, the node betweenness centrality in the case of all single nodes, and edge betweenness centrality in the case of the edges.
\begin{figure*}
    \centering
    \begin{subfigure}{0.33\textwidth}
      \centering
      \includegraphics[width=1\textwidth]{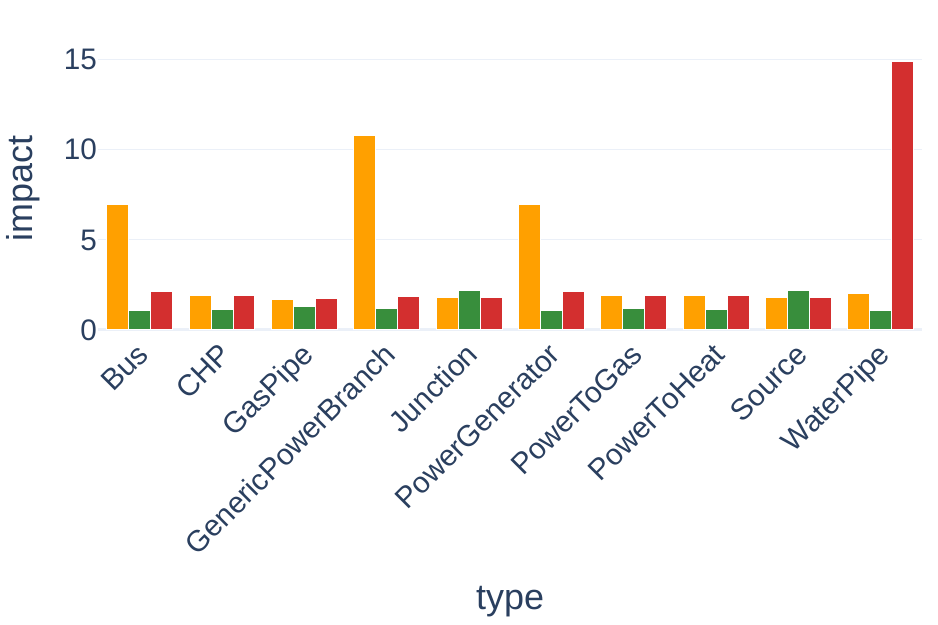}
      \subcaption{Average impacts by component type }\label{fig:avg_impact_component}
    \end{subfigure}
    \begin{subfigure}{0.33\textwidth}
      \centering
      \includegraphics[width=1\textwidth]{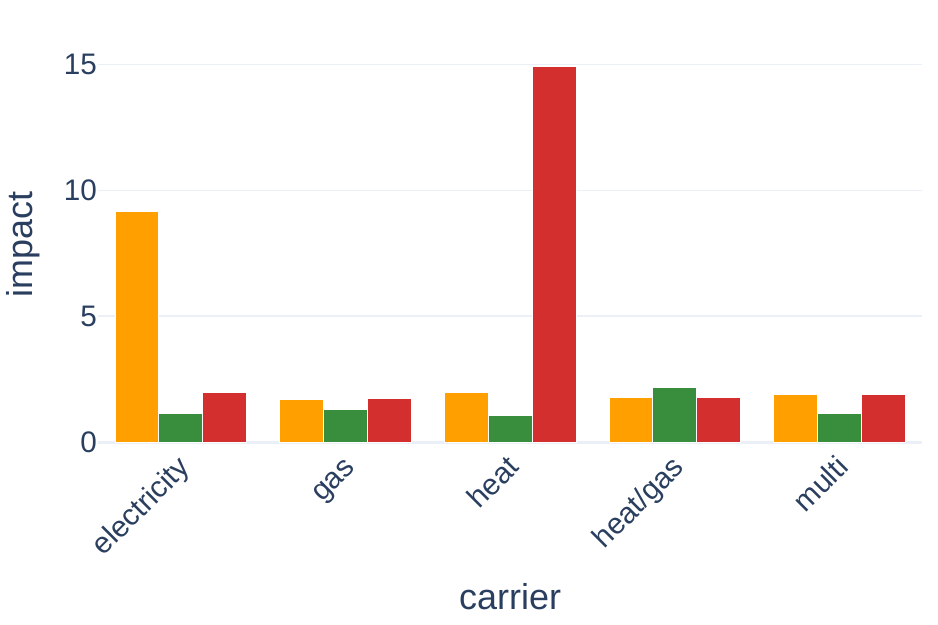}
      \subcaption{Average impacts by carrier type}\label{fig:avg_impact_carrier}
    \end{subfigure}
    \begin{subfigure}{0.33\textwidth}
      \centering
      \includegraphics[width=1\textwidth]{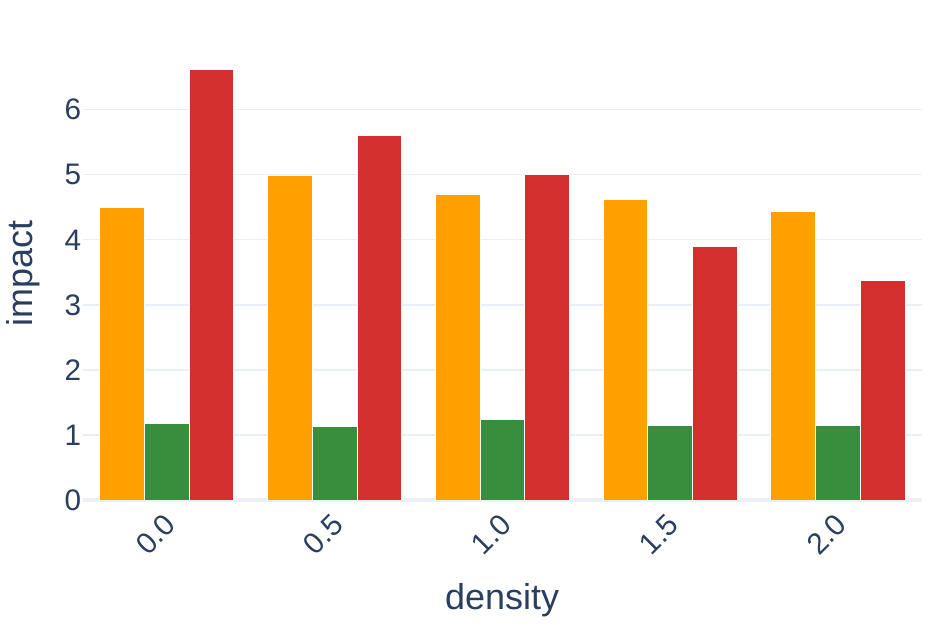}
      \subcaption{Average impacts by density}\label{fig:avg_impact_density}
    \end{subfigure}
    \begin{subfigure}{0.33\textwidth}
      \centering
      \includegraphics[width=1\textwidth]{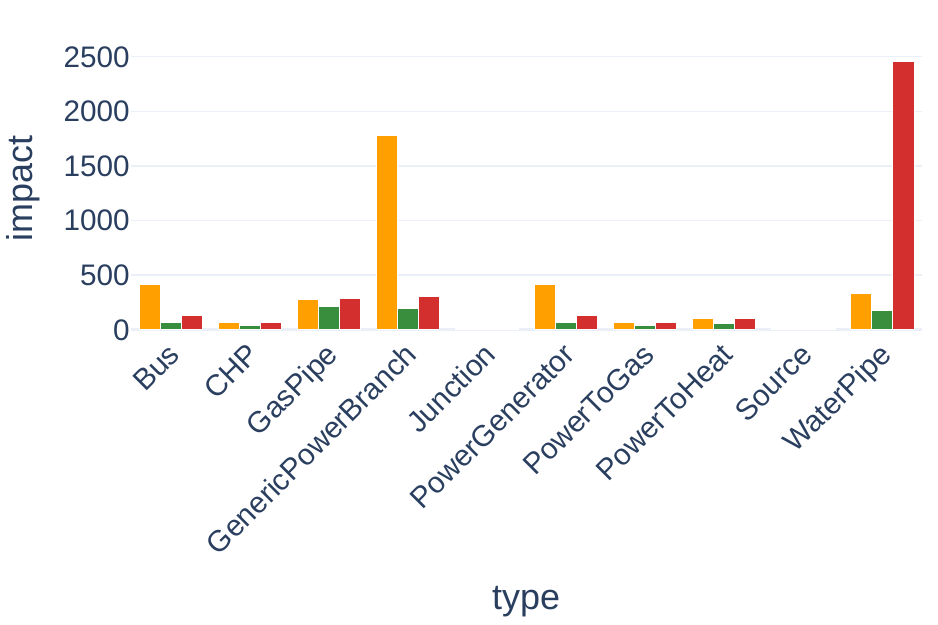}
      \subcaption{Sum of the impacts by component type}\label{fig:total_impact_component}
    \end{subfigure}
    \begin{subfigure}{0.33\textwidth}
      \centering
      \includegraphics[width=1\textwidth]{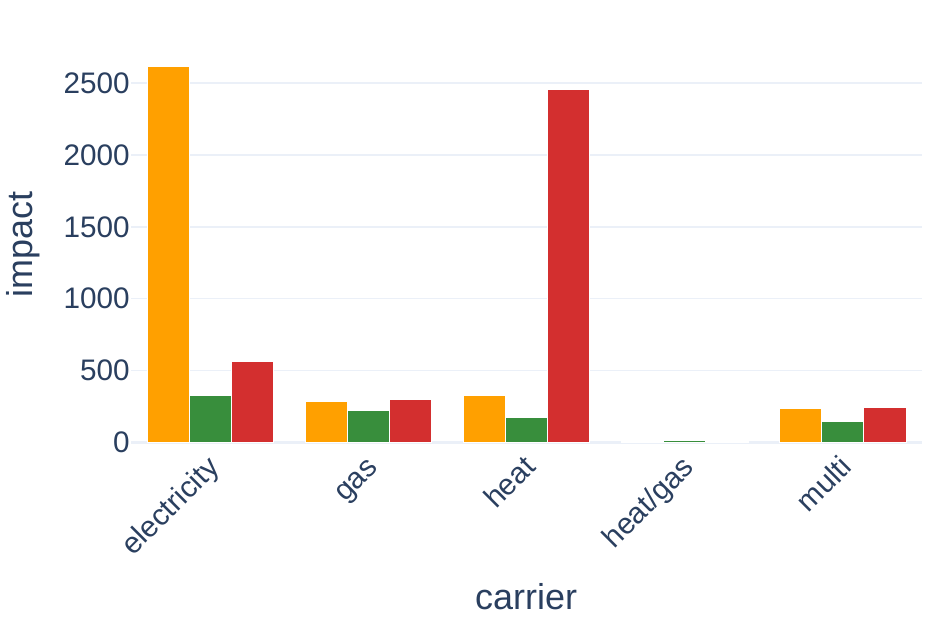}
      \subcaption{Sum of the impacts by carrier type}\label{fig:total_impact_carrier}
    \end{subfigure}
    \begin{subfigure}{0.33\textwidth}
      \centering
      \includegraphics[width=1\textwidth]{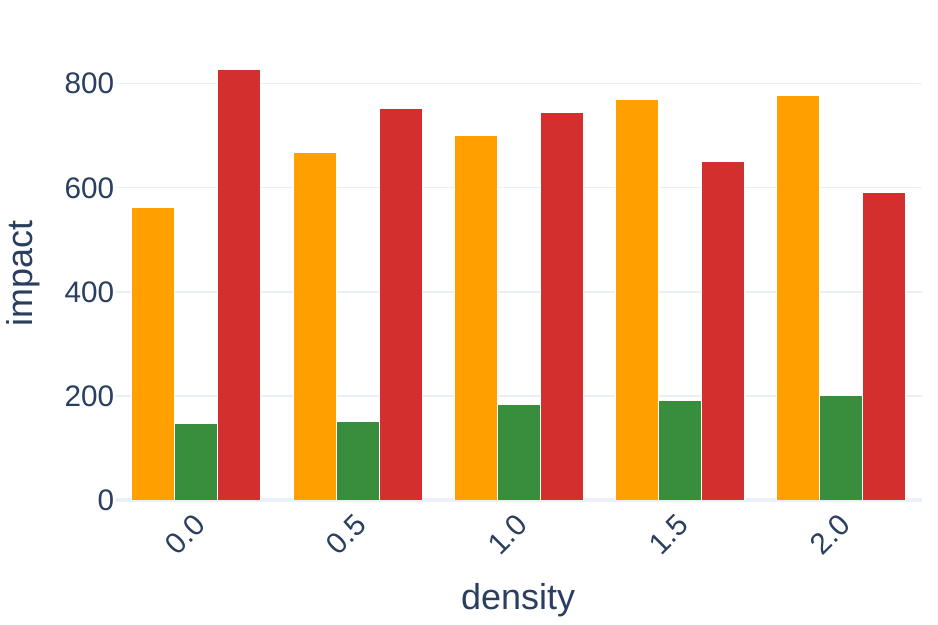}
      \subcaption{Sum of the impacts by density}\label{fig:total_impact_density}
    \end{subfigure}
    \caption{Impacts on the different carrier grids -- \fcolorbox{lightgray}{elcolor}{\rule{0pt}{5pt}\rule{3pt}{0pt}} electricity\space\space\fcolorbox{lightgray}{heatcolor}{\rule{0pt}{5pt}\rule{3pt}{0pt}} heating\space\space\fcolorbox{lightgray}{gascolor}{\rule{0pt}{5pt}\rule{3pt}{0pt}} gas}\label{fig:impact_component}
\end{figure*}
\section{Results}
In this section, we present the results of our experiments. This section is divided into four subsections, each dedicated to answering one or multiple research questions. Firstly, we will evaluate the aggregated resilience of the system by carrier grid. Secondly, we will demonstrate the influence of the density of \ac{CPs} in the system. Thirdly, we will plot the resilience impact by carrier and component from two different perspectives: aggregated and individual. Lastly, we will show the relationship between the impact metric and the topological attributes of the nodes and edges.
\subsection{Resilience by carrier grid}
The first goal is validating the metrics and the Monte Carlo approach. We use the different impact parameter sets and plot the system resilience by scenario and sector. The resilience (as a loss) is calculated according to equation (\ref{eq:resilience_overall}). The result is plotted in Figure \ref{fig:perf_drop_carrier}. The numbers on the x-axis are the CP densities with all parameter-sets in each of their group.

There is a tendency for performance drops to increase with the introduction of the event bias towards the single sectors. For example, a high impact on the electricity sector leads to high values for ${R^G}_LS$. It is also visible that the gas network is nearly unaffected by higher event impacts, implying high over-capacities in the grid. It is also interesting that introducing failures of \ac{CPs} leads to lower R-values, especially if there are also failures in other sectors. This might be a sign that the gas network is generally affected by the failure of other sectors if the \ac{CPs} work without restrictions. It makes sense because the gas network predominantly generates heat and power (and \ac{CPs} also count partially as load. Further, it is interesting that high gas failure rates do not lead to high-performance losses in the gas sectors. We assume that this is a cross-over effect with the decreased demand of \ac{CPs} when pipes fail, as \ac{CPs} (\ac{CHP}s in this case) do not count towards the performance loss in general.
\subsection{Influence of the CP density}
One crucial aspect is the evaluation of different CP capacity installments. For this, five different CP densities were tested, and for every experiment, the performance drop is shown in Figure \ref{fig:perf_drop_carrier}. Further, we evaluate the influence of CP densities on the impact metric, shown in Figures \ref{fig:avg_impact_density} and \ref{fig:total_impact_density}.

In Figure \ref{fig:perf_drop_carrier}, the influence of the different CP densities is high. Two clear relations are shown: if the density is higher, the gas sector's performance will decrease (remember that it is always about emergencies), while the electricity sector's performance is higher. The heating sectors' performance increases significantly when using a higher CP density. Generally, we can expect the heating sector to profit from \ac{CPs}, as this sector only receives additional energy. 

Figures \ref{fig:avg_impact_density} and \ref{fig:total_impact_carrier} show that the overall aggregated average will decrease with higher densities, which implies that higher CP-densities increase the resilience of the overall system. We conclude that the CP density has a slightly positive influence on the overall resilience, achieved by using the gas systems' capabilities to balance the overall performance. This result seems valid as we look at overall performance when calculating the optimal load shedding. 
\begin{figure*}
    \centering
    \begin{subfigure}{0.33\textwidth}
      \centering
      \includegraphics[width=1\textwidth]{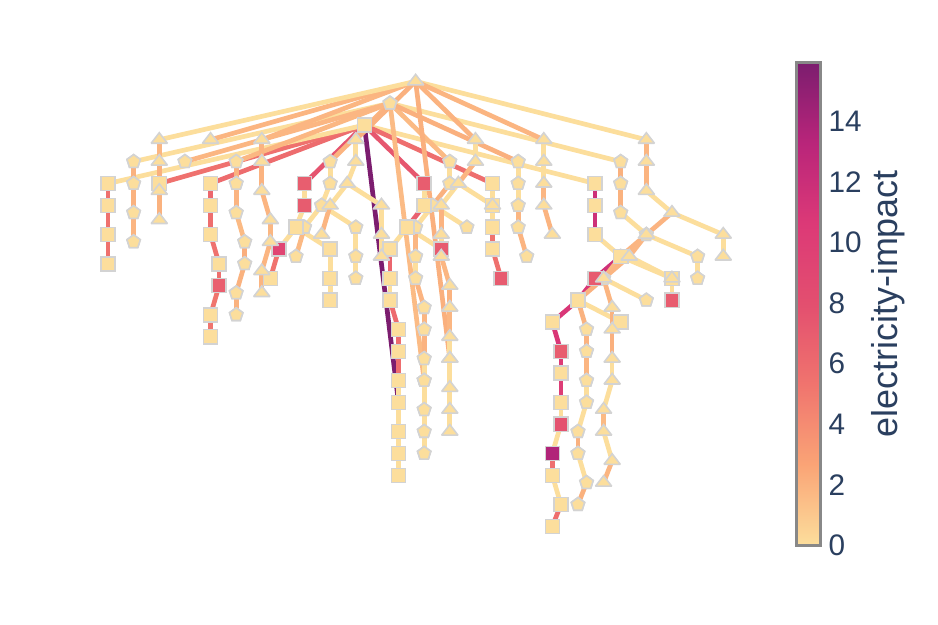}
      \subcaption{Components' total impacts in the network with a CP-density of 0.5}\label{fig:graph_impact_05}
    \end{subfigure}
    \begin{subfigure}{0.33\textwidth}
      \centering
      \includegraphics[width=1\textwidth]{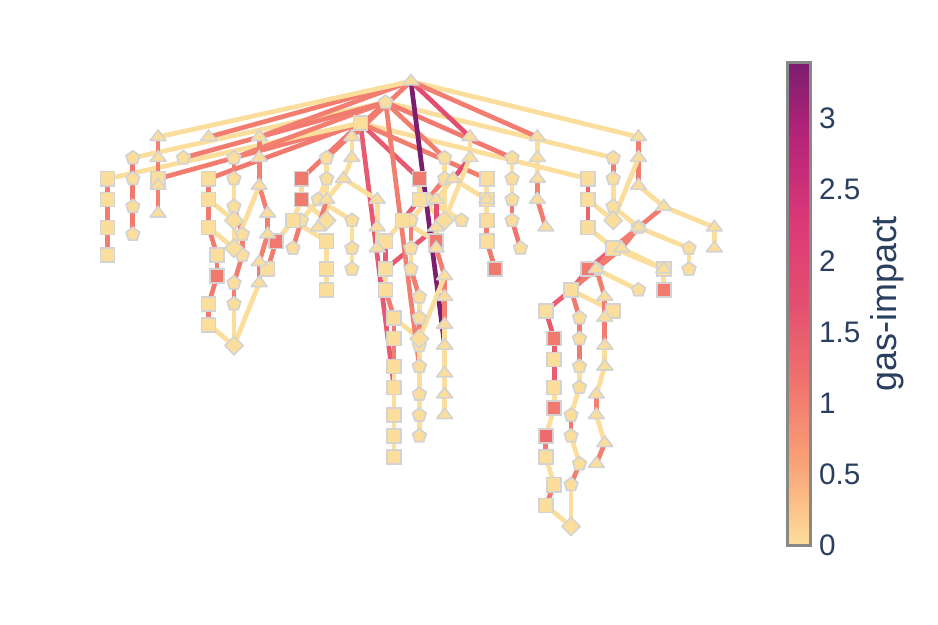}
      \subcaption{Branches' gas impacts in the network with a CP-density of 0.5}\label{fig:graph_impact_gas_05}
    \end{subfigure}
    \begin{subfigure}{0.33\textwidth}
      \centering
      \includegraphics[width=1\textwidth]{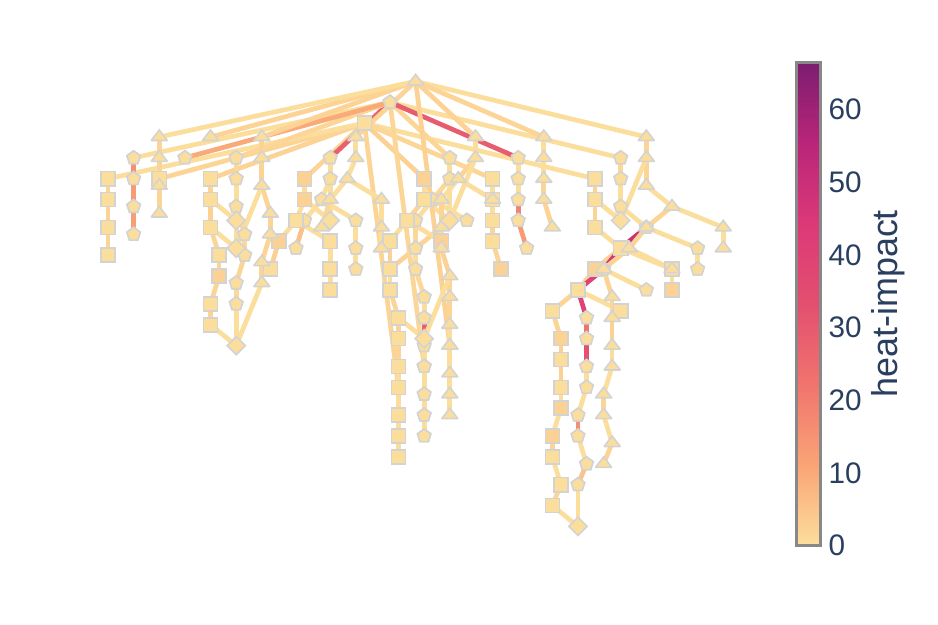}
      \subcaption{Branches' heat impacts in the network with a CP-density of 0.5}\label{fig:graph_impact_heat_05}
    \end{subfigure}
    \begin{subfigure}{0.33\textwidth}
      \centering
      \includegraphics[width=1\textwidth]{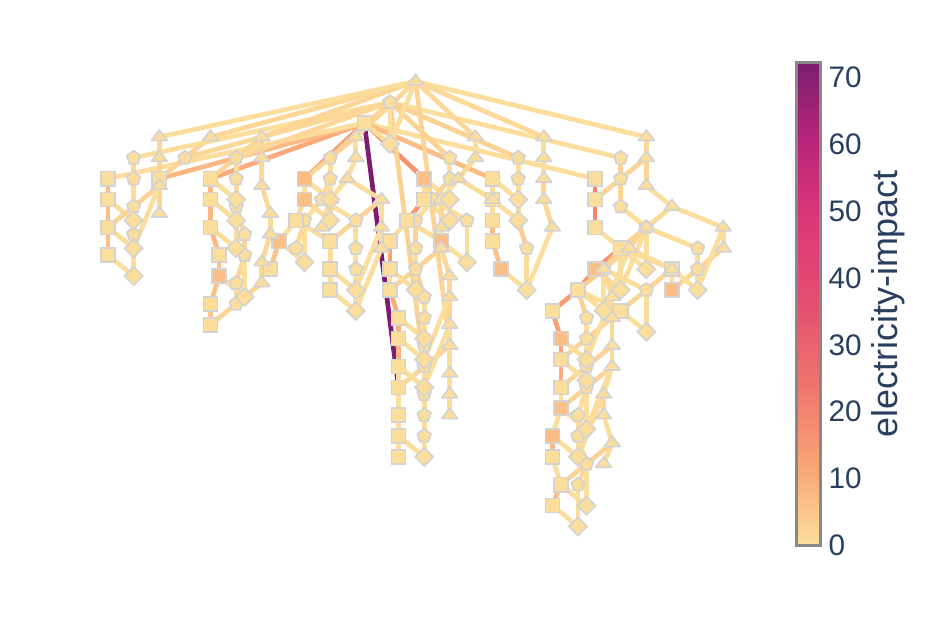}
      \subcaption{Components' total impacts in the network with a CP-density of 2}\label{fig:graph_impact_20}
    \end{subfigure}
    \begin{subfigure}{0.33\textwidth}
      \centering
      \includegraphics[width=1\textwidth]{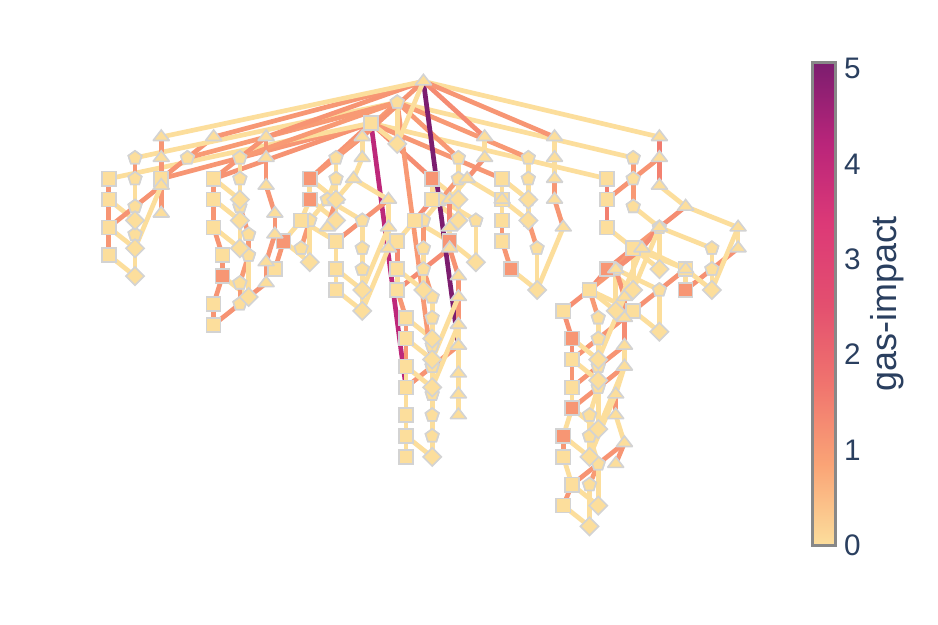}
      \subcaption{Branches' gas impacts in the network with a CP-density of 2}\label{fig:graph_impact_gas_20}
    \end{subfigure}
    \begin{subfigure}{0.33\textwidth}
      \centering
      \includegraphics[width=1\textwidth]{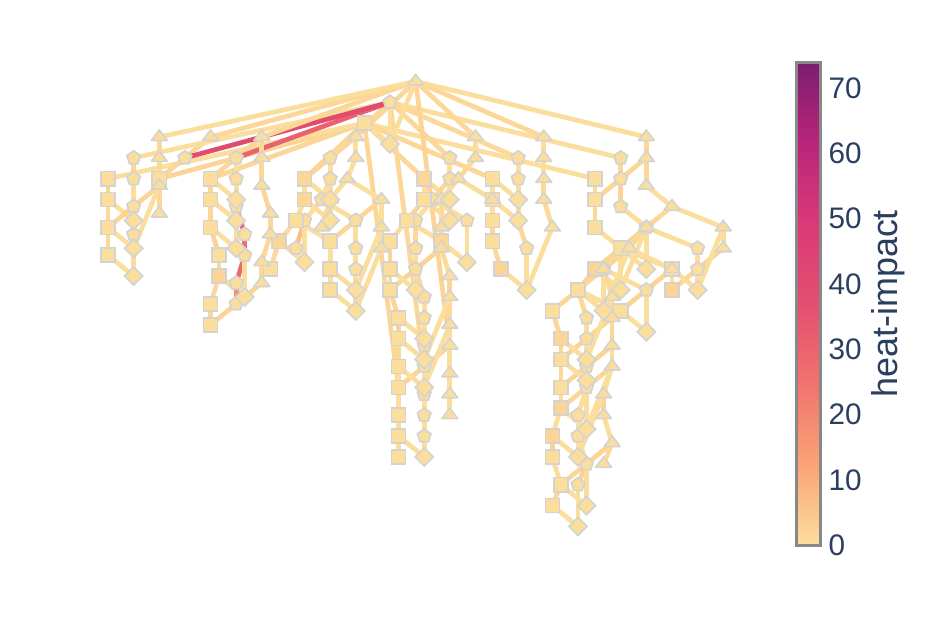}
      \subcaption{Branches' heat impacts in the network with a CP-density of 2}\label{fig:graph_impact_heat_20}
    \end{subfigure}
    \caption{Network visualization of the multi-grid system with different carrier-impacts of the components as main-color; the carrier of the nodes are depicted as the shape (rectangle = electricity, triangle = gas, pentagon = heat, cp = diamond}\label{fig:graph_impact_results}
\end{figure*}
\subsection{Resilience Impact}
In this section, the impact metric, aggregated by component type, carrier, and individual, is calculated (according to equation (\ref{eq:impact_ind}) and (\ref{eq:impact_grid})) and plotted. The average impact, aggregated by component type and segmented by the carrier, is depicted in Figure \ref{fig:avg_impact_component}, and the total impact is depicted in Figure \ref{fig:total_impact_component}. Further, we show the average impact of the carriers on each other (Figure \ref{fig:avg_impact_carrier}) and the total impact of the carriers on each other (Figure \ref{fig:total_impact_carrier}).

Most values are reasonable. For example, the impact of the water pipe in the heating grid mainly affects the heating, which makes sense, as there is no energy flowing from the heating grid in the other energy carrier grids. Further, it is notable that the impact carrier grids are generally low for \ac{CPs}. Failure of (only) \ac{CPs}, in general, does not affect the grid in a significant way.

Looking at the impacts of the carrier, it is clear that the heating carrier only affects itself while gas affects electricity and heat if \ac{CPs} exist; electricity majorly affects itself, and the heating grid and the \ac{CPs} affect all networks minorly (probably due to the stochastic nature of the simulation).

The second way to visualize the impact is by plotting the network as a node or edge graph. Figures \ref{fig:graph_impact_05} and \ref{fig:graph_impact_20} show the impact for every node and edge. The Figures \ref{fig:graph_impact_gas_05} and \ref{fig:graph_impact_gas_20} show the impact on the gas grid. Further, \ref{fig:graph_impact_heat_05} and \ref{fig:graph_impact_heat_20} depict the impact on the heating grid.

These visualizations show that, first, the high impacts generally are driven by branches. There is a strong bias of the impact towards more central edges, especially in highly populated areas (connected to high-degree nodes). It further makes clear that the highest impact on each carrier is still the carrier itself. When looking at the differences in the impacts of different CP-density networks, the lower overall impact is shown in less central areas. 
\subsection{Relation to topological attributes}
\begin{figure*}
    \centering
    \begin{subfigure}{0.32\textwidth}
      \centering
      \includegraphics[width=1\textwidth]{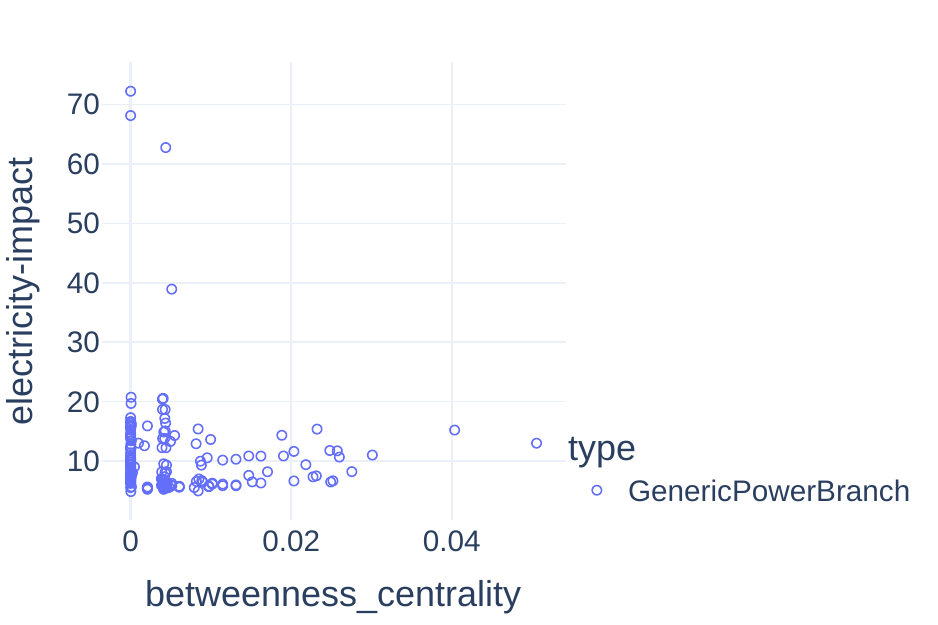}
      \subcaption{Betweenness centrality to the impact on the electricity grid of the electric lines}\label{fig:impact_bc_el}
    \end{subfigure}
    \begin{subfigure}{0.32\textwidth}
      \centering
      \includegraphics[width=1\textwidth]{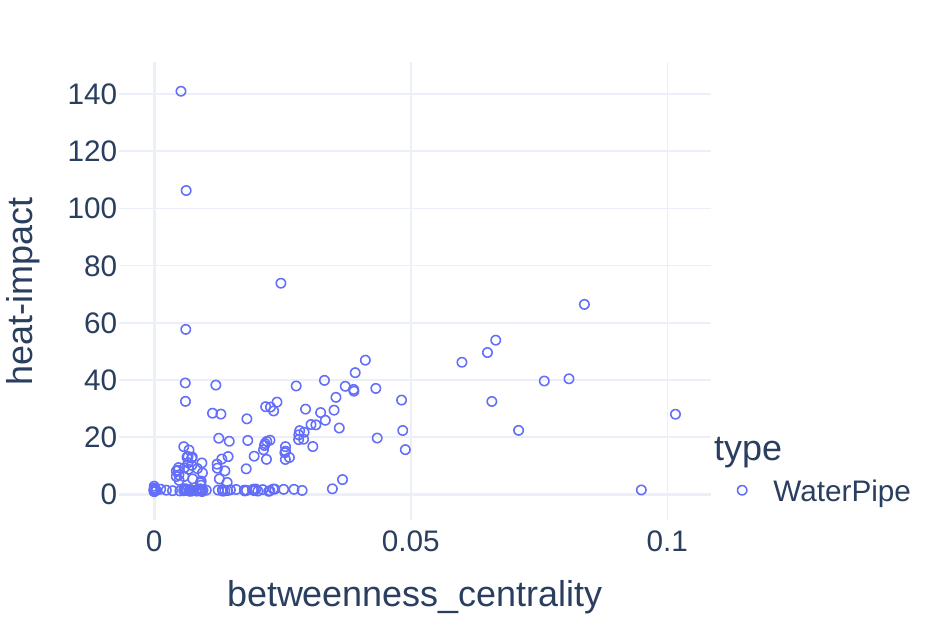}
      \subcaption{Betweenness centrality to the impact on the heating grid of the water pipes}\label{fig:impact_bc_heat}
    \end{subfigure}
    \begin{subfigure}{0.32\textwidth}
      \centering
      \includegraphics[width=1\textwidth]{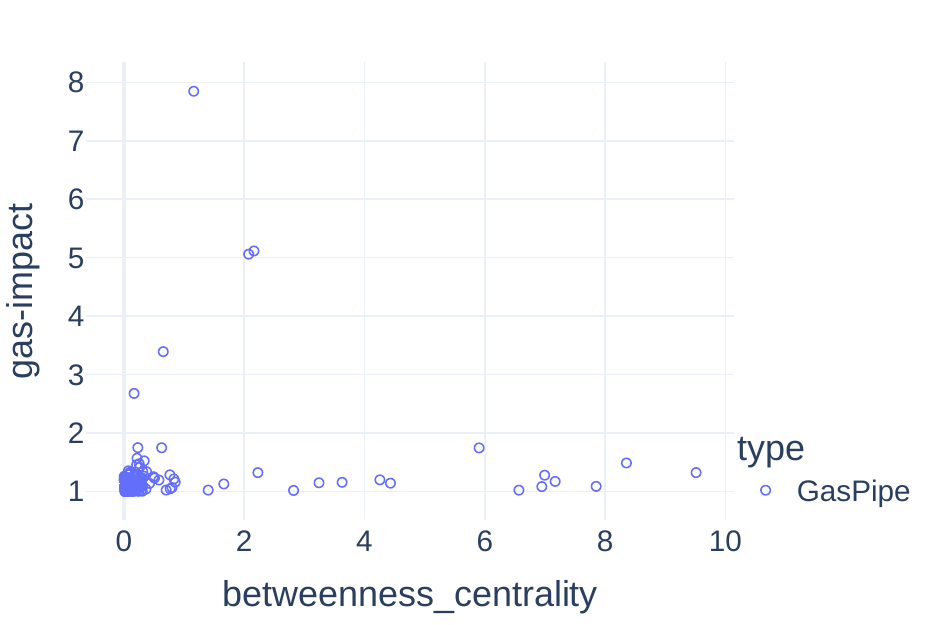}
      \subcaption{Betweenness centrality to the impact on the gas grid of the gas pipes}\label{fig:impact_bc_gas}
    \end{subfigure}
    \begin{subfigure}{0.32\textwidth}
      \centering
      \includegraphics[width=1\textwidth]{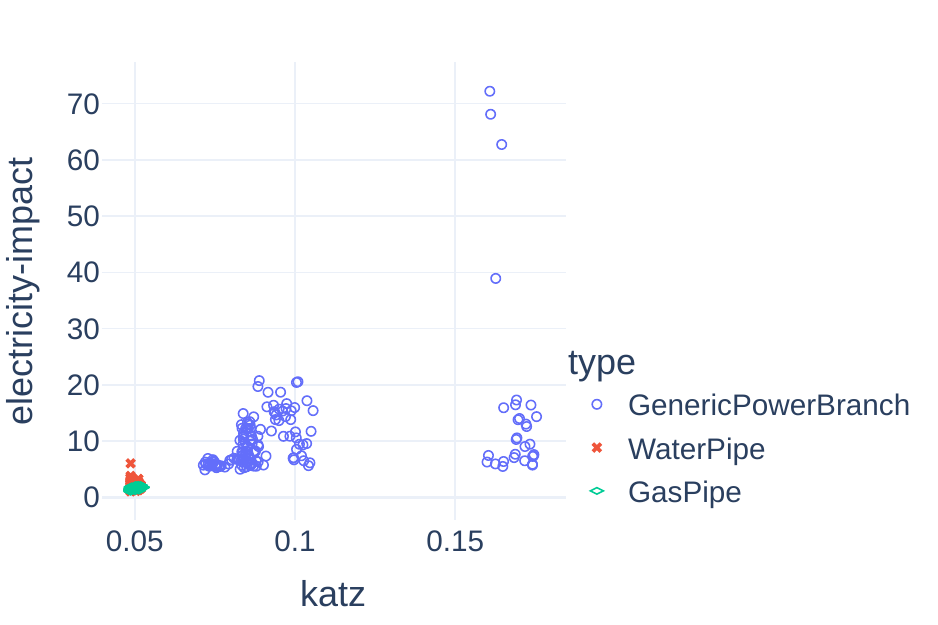}
      \subcaption{Katz centrality to the impact on the electricity grid of the branches}\label{fig:impact_katz_el}
    \end{subfigure}
    \begin{subfigure}{0.32\textwidth}
      \centering
      \includegraphics[width=1\textwidth]{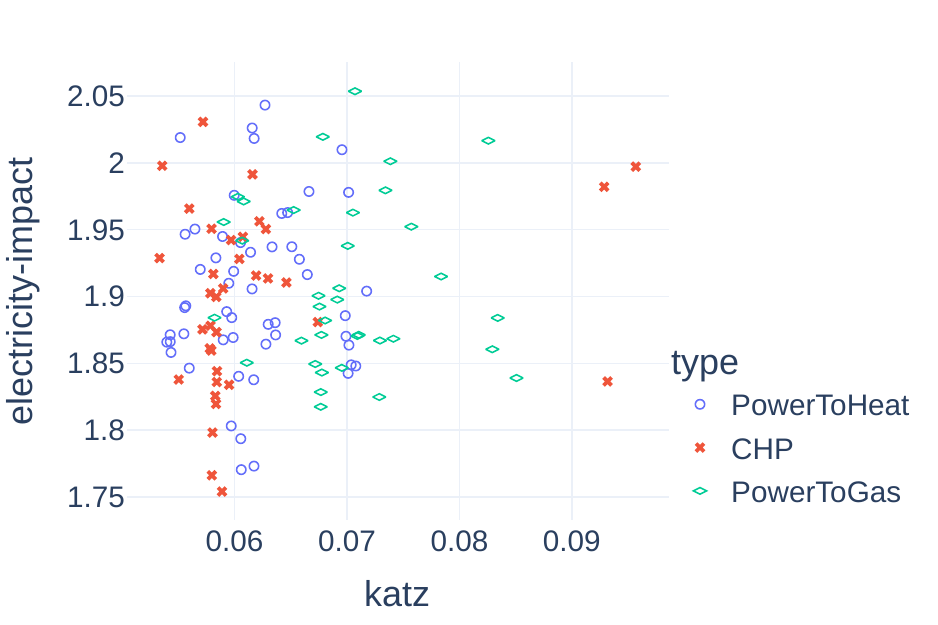}
      \subcaption{Katz centrality to the impact on the electricity grid of the CPs}\label{fig:impact_katz_heat}
    \end{subfigure}
    \begin{subfigure}{0.32\textwidth}
      \centering
      \includegraphics[width=1\textwidth]{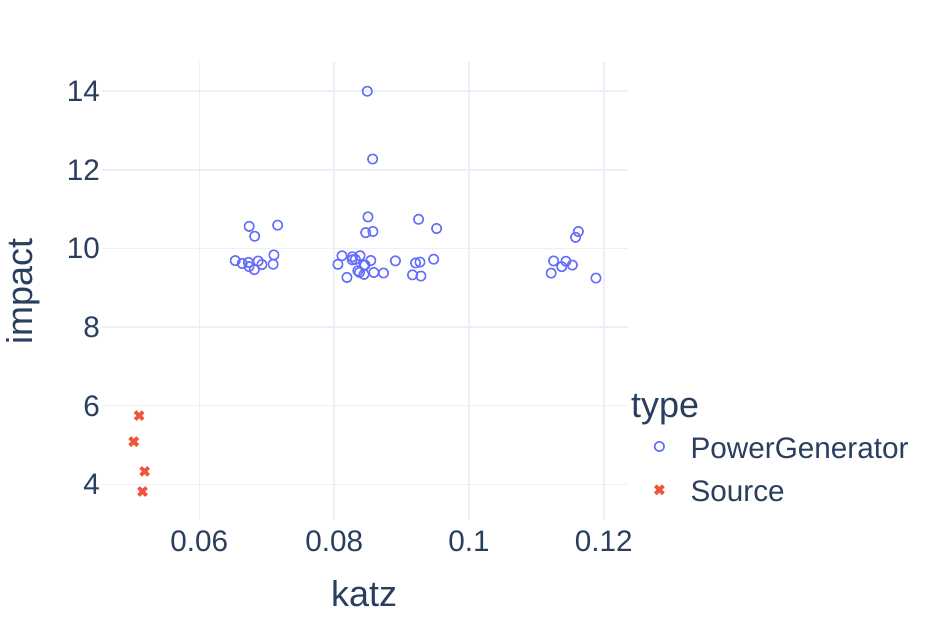}
      \subcaption{Katz centrality to the overall impact of the power generators}\label{fig:impact_katz_gas}
    \end{subfigure}
    \caption{Relation between graph metrics and the impact of the components on the resilience}\label{fig:res_metric_relations}
\end{figure*}

At this point, we validated the effectiveness of the impact metric and now aim to find correlations between the topological attributes of the nodes and edges and the impact metric. We plot the impact of every single node and edge to topological metrics. The betweenness centrality of the branches to the electricity, gas, and heat branches are depicted in Figures \ref{fig:impact_bc_el}, \ref{fig:impact_bc_heat}, and \ref{fig:impact_bc_gas}. Further, the Katz centrality of several component types (\ac{CPs}, branches, nodes) to the impact on the branches, \ac{CPs} and generators are depicted in Figures \ref{fig:impact_katz_el}, \ref{fig:impact_katz_gas}, and \ref{fig:impact_katz_heat}.

Looking at most of the components, the metrics cannot strongly capture the resilience impact behavior. However, especially when looking at the branches, there is a clear correlation using the Katz centrality (and betweenness centrality), which shows that the main impact factor, topologically wise, is the degree of the nodes and their neighbor's degrees.

We note that the heating CPs (\ac{P2H} and \ac{CHP}) also show a slight correlation when looking at the KC. However, in general, the overall impact of CPs on the system has to be higher, such that their single impact and correlation are significant.

There is no clear correlation between the centrality metrics and any node impacts.  
\subsection{Conclusion}
At last, we will conclude the results and directly answer the research questions raised initially.
\paragraph{Do the grids influence each other's resilience, and if so, what is the impact of the coupling density on this effect?}
The carrier grids influence each other, especially when looking at the relationship between electricity and the gas grid, in which the influences change majorly when increasing the coupling density/capabilities.
\paragraph{Can couplings between the grids increase or decrease resilience (considering no specific island-building strategies)?}
It increases the resilience using the model presented, which is limited in terms of being able to operate the islanded parts of the single carrier grids. Further, this increase demands overcapacities in the gas grid (we assume it would also work the other way around). The related work indicates that using \ac{CPs} and islanding further increases resilience. Further, it can decrease the resilience of single-carrier grids, as seen with the gas grid in this case study. However, it will never decrease the resilience of the system as a whole as long as the optimization problem/operating strategy reflects the type of resilience that will be achieved. Note that this implies that not optimal operation strategies can decrease the system's resilience because \ac{CPs} generally have a lower efficiency than regular branches.
\paragraph{Is there a relation between complex topology attributes and the resilience/resilience impact?}
There is a relation in some instances. The impact of electricity grids' branches and the \ac{P2H} and \ac{CHP} components show a relation to the Katz centrality, degree, and betweenness centrality. However, the impacts of most components are driven by their capabilities. However, enhancing topological metrics using the components' physical attributes might be possible.  
\section{Summary and Outlook}
This paper presents a novel approach to generating high-impact events and simulating the events to calculate defined resilience metrics in coupled energy grids. The single carrier grids are physically modeled using the steady-state equations. An approach is shown to convert the grid to a graph representation. Further, we define the impact metric, which can capture the influence of nodes and edges on the carrier grids' resilience. After that, the Monte Carlo simulation is introduced, the experiments are defined, and the results are shown. 

First, the results show that the simulation is feasible for calculating the network's resilience. Second, the impact metric can describe the influence of the nodes and edges. Further, there is a relationship between topological attributes and the impact of resilience. However, there is no indication of a general correlation.

In the future, the model and simulation can be improved on multiple levels. One can use other grid topologies, especially topologies with cycles. Further, the data can be improved. We used static power values, which can be substituted with time series data. Other than this, the model quality, especially the heating grid, can be improved. For example, the way the heating introduction grid is modeled makes it impossible to introduce cycles, and it is not feasible to use return pipes. Further, the simulation of our scenarios needs much computational time due to the number of optimization problems that need to be solved. In the future, it might be beneficial, especially if one wants to analyze multiple different coupled grids, to approximate the load-shedding optimization by using deep learning approaches to significantly lower the simulation time and, therefore, increase the number of feasible development iterations, which could also further increase the quality of the results. 
\section*{Acknowledgments}

This work has been funded by the Deutsche Forschungsgemeinschaft (DFG, German Research Foundation) – 359941476.

The simulations were performed at the HPC Cluster ROSA, located at the University of Oldenburg (Germany) and funded by the DFG through its Major Research Instrumentation Programme (INST 184/225-1 FUGG) and the Ministry of Science and Culture (MWK) of the Lower Saxony State.

\bibliographystyle{ieeetr}
\bibliography{main}

\appendix

\section{AC Power Equations}\label{app:ac_equations}
We conveniently insert the well-known AC steady-state power as they are formulated by PowerModels \cite{8442948} (shortened version) in the following section.
\begin{equation}
\begin{split}
& S^g_k \;\; \forall k\in G \\
& S^d_k \;\; \forall k\in D \\
& V_i \;\; \forall i\in N \\
& S_{ij} \;\; \forall (i,j) \in E \\
\end{split}
\end{equation}

\begin{equation}
\begin{split}
& \sum_{\substack{k \in G_i}} S^g_k - \sum_{\substack{k \in L_i}} S^d_k - \sum_{\substack{k \in S_i}} (Y^s_k)^* |V_i|^2 = \sum_{\substack{(i,j)\in E_i \cup E_i^R}} S_{ij} \;\; \forall i\in N \label{eq_kcl_shunt} \\
& S_{ij} = \left( Y_{ij} + Y^c_{ij}\right)^* \frac{|V_i|^2}{|{T}_{ij}|^2} - Y^*_{ij} \frac{V_i V^*_j}{{T}_{ij}} \;\; \forall (i,j)\in E\\
& S_{ji} = \left( Y_{ij} + Y^c_{ji} \right)^* |V_j|^2 - Y^*_{ij} \frac{V^*_i V_j}{{T}^*_{ij}} \;\; \forall (i,j)\in E\\
\end{split}
\end{equation}
Here, $S^g_k$ is the power dispatched by generator $k$, $S^d_k$ is the power dispatched by demand $k$, $G$ is the set of generators, $D$ is the set of demands, $V_i$ is the complex bus voltage of bus $i$, $N$ is the set of buses, $S_{ij}$ is the branch complex power flow between $i$ and $j$, $E$ is the set of branches. Further, $Y_{ij}$ is the admittance between $i$ and $j$, $Y_{ij}^c$ is the directed line conductance, $T_{ij}$ is the transformation rate. 
\section{Network Visualization (no CPs) and Additional Result Figures}
\begin{figure}[hbt]
    \centering
    \includegraphics[width=0.4\textwidth]{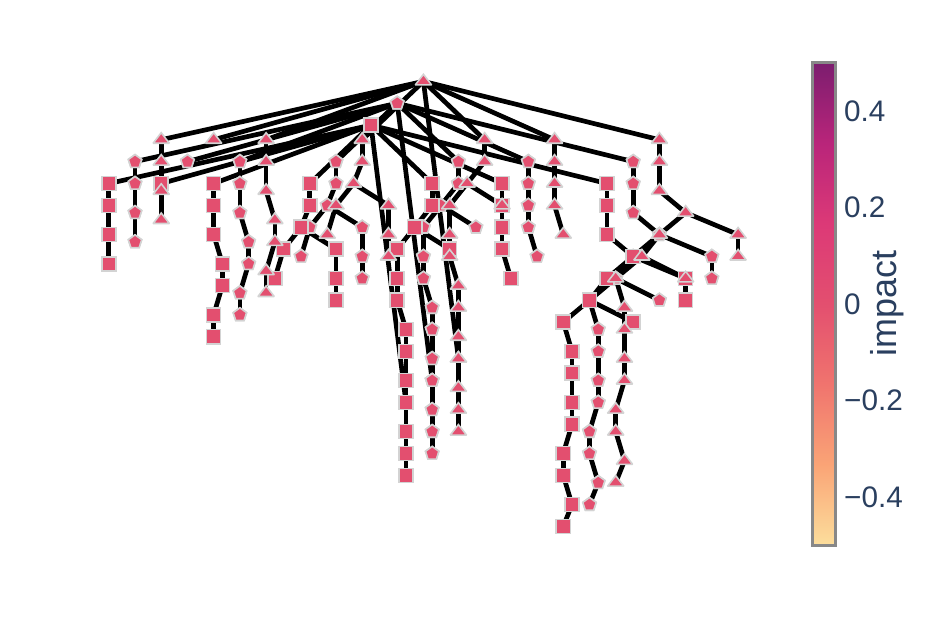}
    \caption{Network visualization of the multi-grid base system (without \ac{CPs}) with impact as main-color; the carrier of the component is depicted as shape (rectangle = electricity, triangle = gas, pentagon = heat, cp = diamond}\label{fig:graph_impact_results_appendix}
\end{figure}
\begin{figure*}[hbt]
    \centering
    \begin{subfigure}{0.32\textwidth}
      \centering
      \includegraphics[width=1\textwidth]{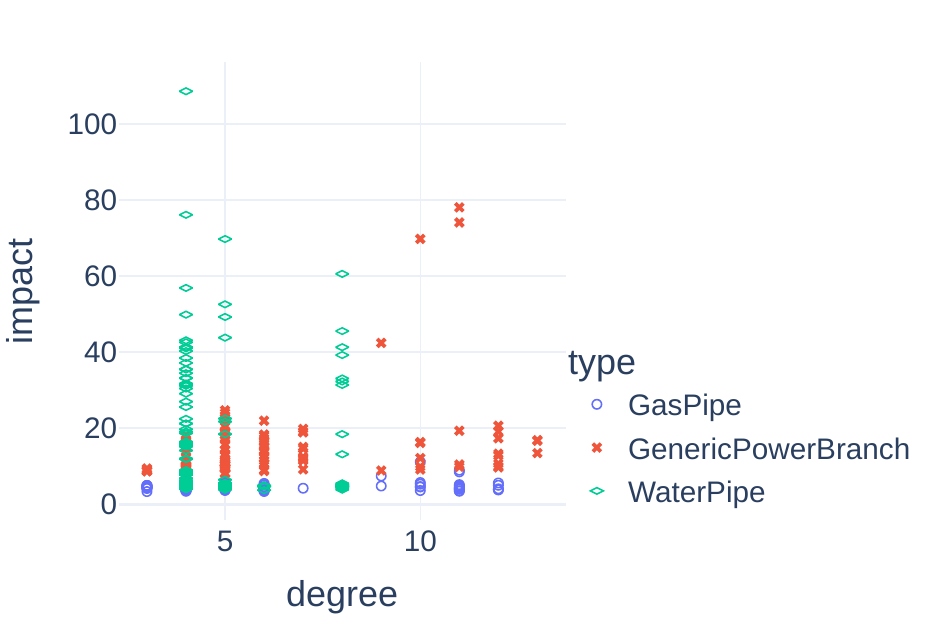}
      \subcaption{Degree to the overall impact of the branches}\label{fig:impact_degree_branches}
    \end{subfigure}
    \begin{subfigure}{0.32\textwidth}
      \centering
      \includegraphics[width=1\textwidth]{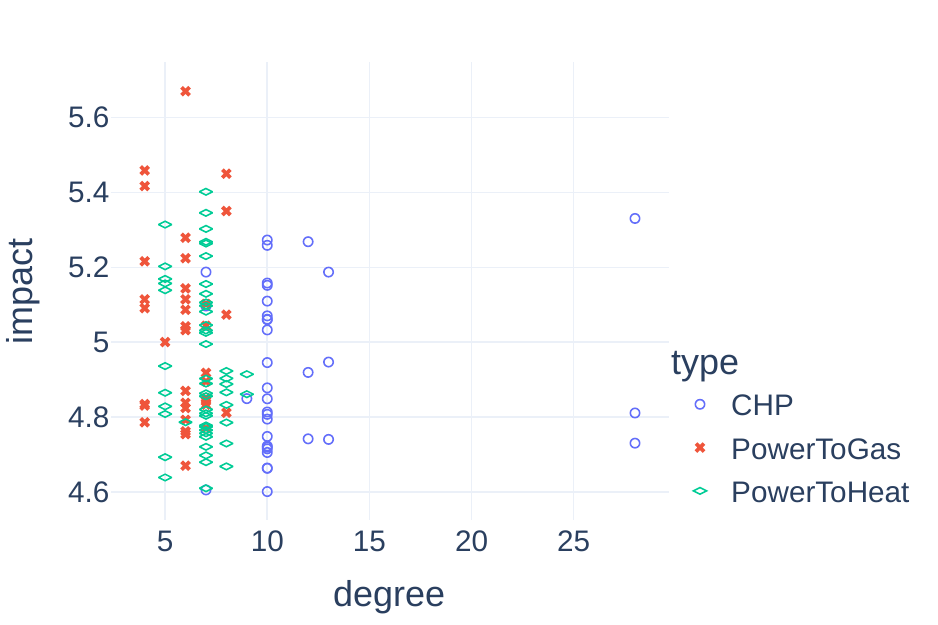}
      \subcaption{Degree to the overall impact of the CPs}\label{fig:impact_degree_cps}
    \end{subfigure}
    \begin{subfigure}{0.32\textwidth}
      \centering
      \includegraphics[width=1\textwidth]{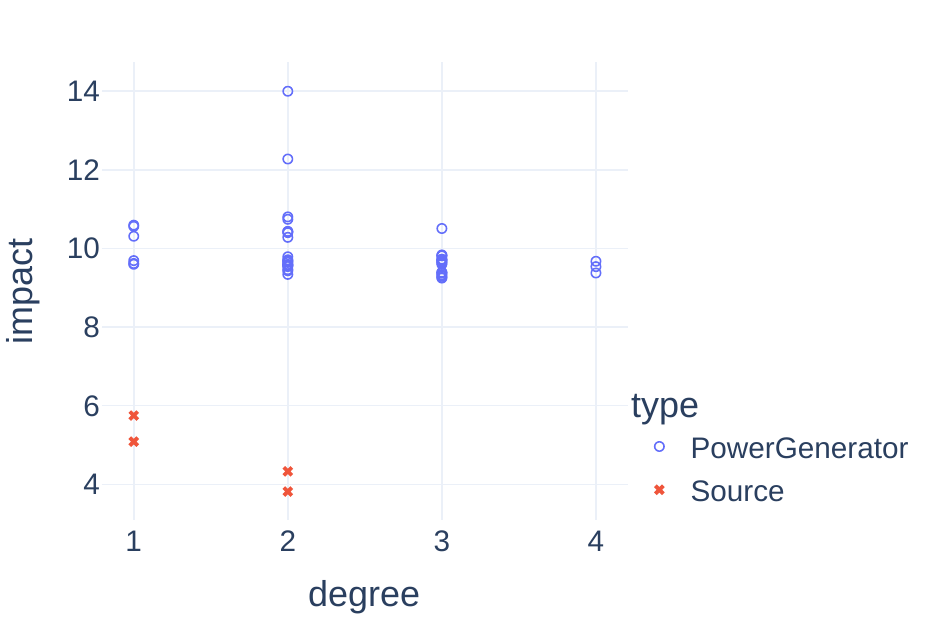}
      \subcaption{Degree to the overall impact of the nodes}\label{fig:impact_degree_nodes}
    \end{subfigure}    
    \begin{subfigure}{0.32\textwidth}
      \centering
      \includegraphics[width=1\textwidth]{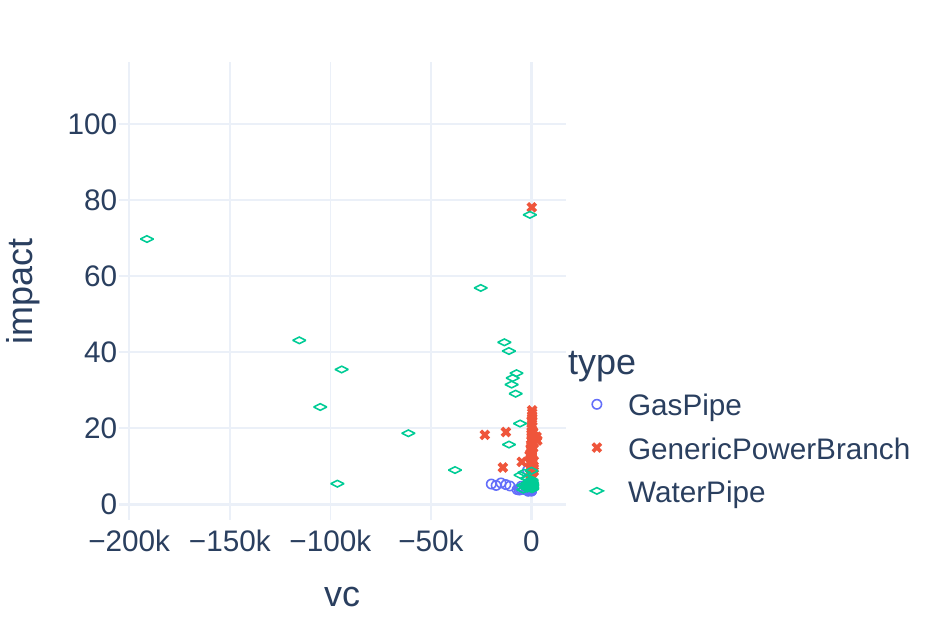}
      \subcaption{Vitality closeness to the overall impact of the branches}\label{fig:impact_vc_branches}
    \end{subfigure}
    \begin{subfigure}{0.32\textwidth}
      \centering
      \includegraphics[width=1\textwidth]{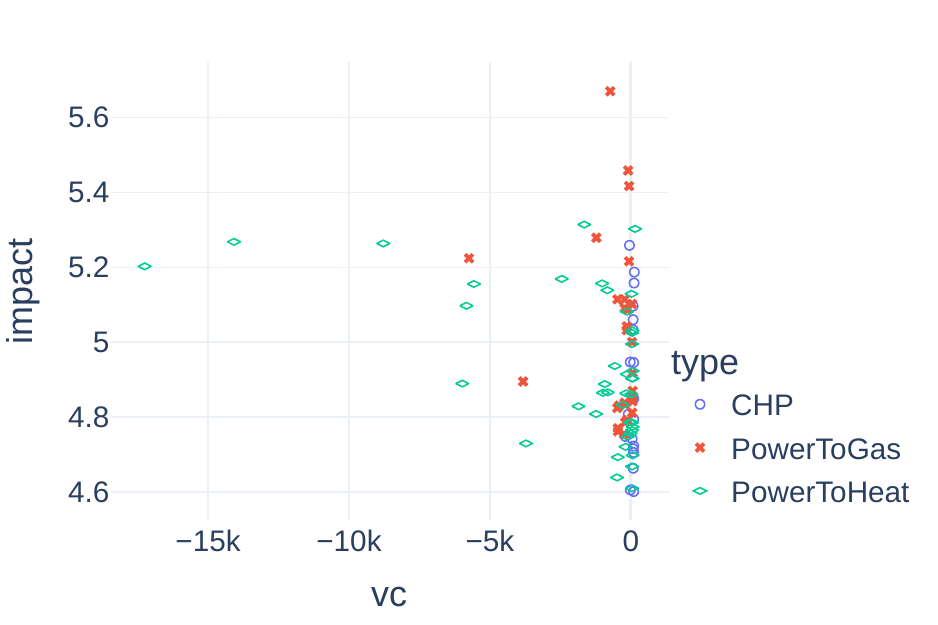}
      \subcaption{Vitality closeness to the overall impact of the CPs}\label{fig:impact_vc_cps}
    \end{subfigure}
    \begin{subfigure}{0.32\textwidth}
      \centering
      \includegraphics[width=1\textwidth]{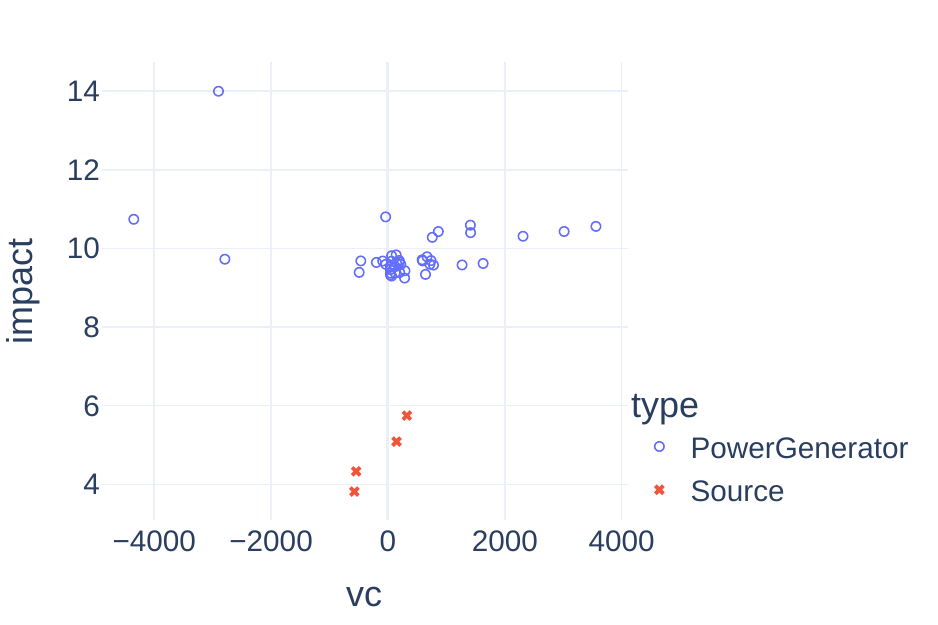}
      \subcaption{Vitality closeness to the overall impact of the nodes}\label{fig:impact_vc_nodes}
    \end{subfigure}
    \begin{subfigure}{0.32\textwidth}
      \centering
      \includegraphics[width=1\textwidth]{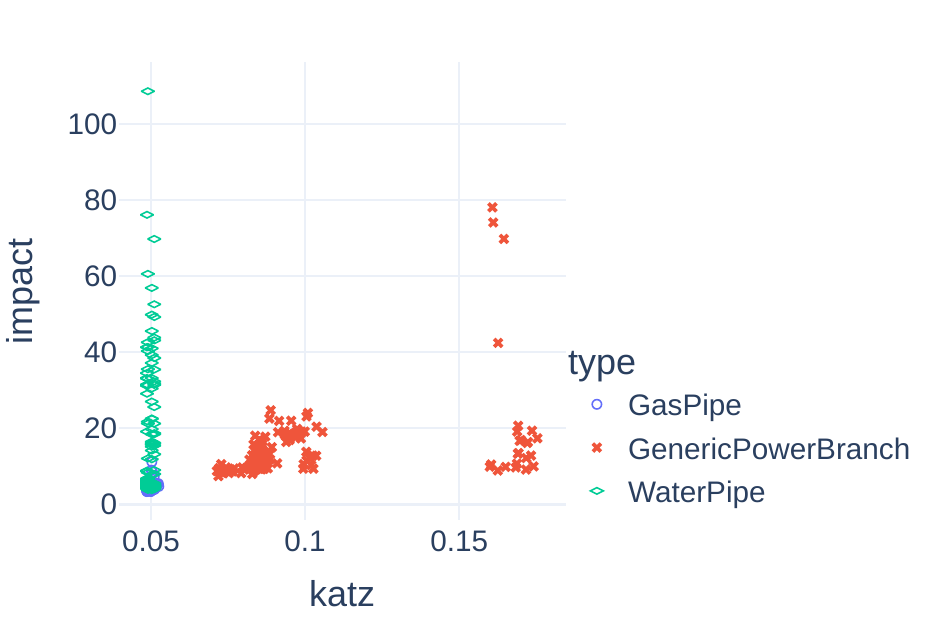}
      \subcaption{Katz centrality to the overall impact of the branches}\label{fig:impact_katz_branches}
    \end{subfigure}
    \begin{subfigure}{0.32\textwidth}
      \centering
      \includegraphics[width=1\textwidth]{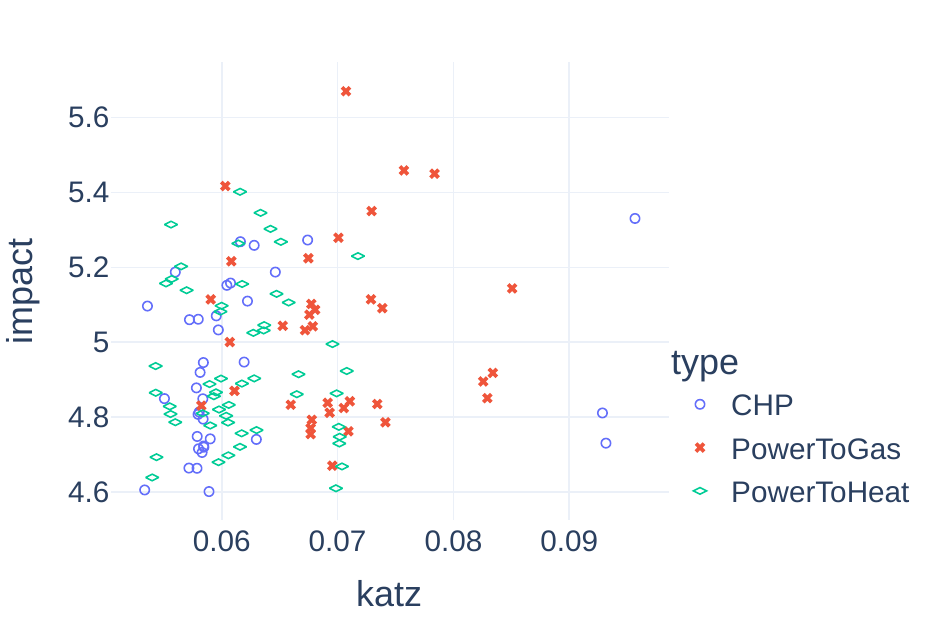}
      \subcaption{Katz centrality to the overall impact of the CPs}\label{fig:impact_katz_cps}
    \end{subfigure}
    \begin{subfigure}{0.32\textwidth}
      \centering
      \includegraphics[width=1\textwidth]{figure/katz_to_the_nodes_heat-impact-katz-impact.pdf}
      \subcaption{Katz centrality to the overall impact of the nodes}\label{fig:impact_katz_nodes}
    \end{subfigure}
    \begin{subfigure}{0.32\textwidth}
      \centering
      \includegraphics[width=1\textwidth]{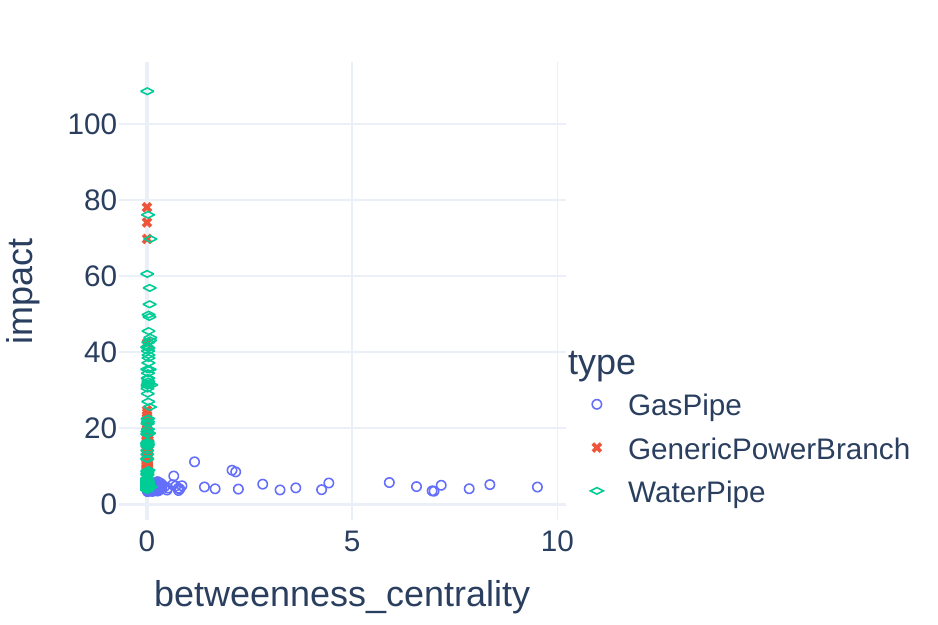}
      \subcaption{Betweenness centrality to the overall impact of the branches}\label{fig:impact_bc_branches}
    \end{subfigure}
    \begin{subfigure}{0.32\textwidth}
      \centering
      \includegraphics[width=1\textwidth]{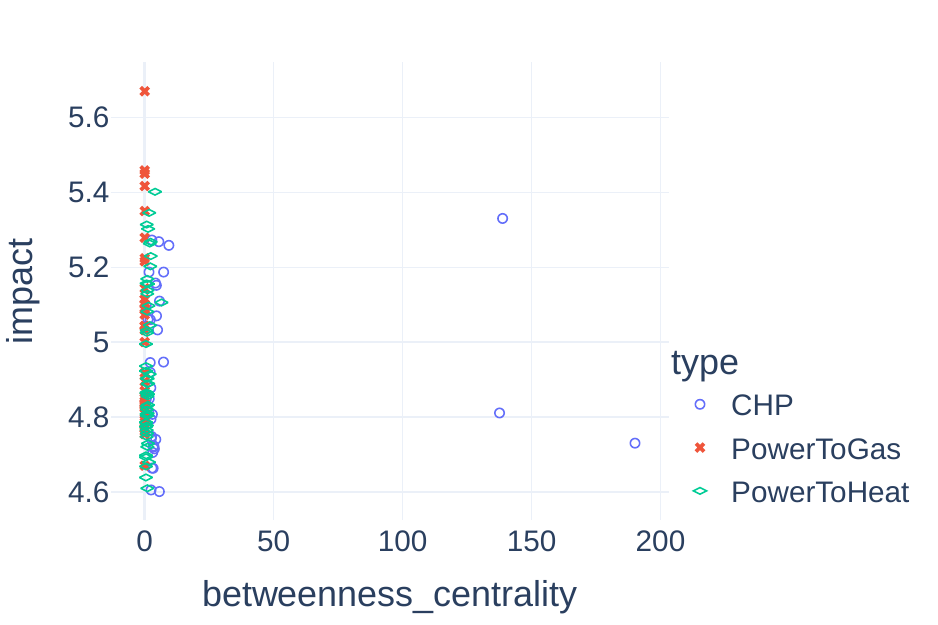}
      \subcaption{Betweenness centrality to the overall impact of the CPs}\label{fig:impact_bc_cps}
    \end{subfigure}
    \begin{subfigure}{0.32\textwidth}
      \centering
      \includegraphics[width=1\textwidth]{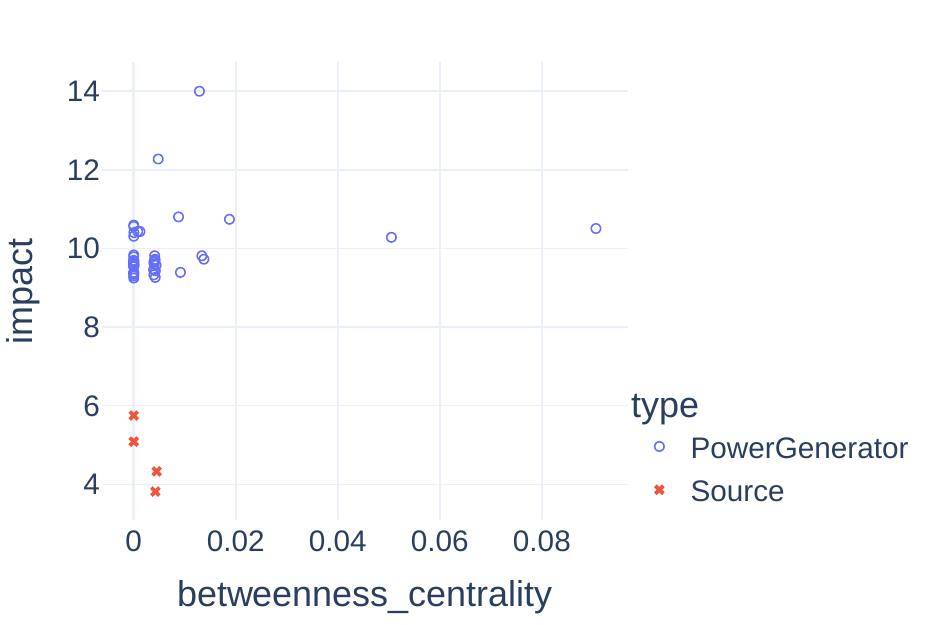}
      \subcaption{Betweenness centrality to the overall impact of the nodes}\label{fig:impact_bc_nodes}
    \end{subfigure}
    
    \caption{Relation between vc and degree metrics and the impact of the components on the resilience}\label{fig:res_metric_relations_appendix}
\end{figure*}

\end{document}